\documentclass[12pt,prb,letterpaper,floatfix,preprintnumbers,superscriptaddress]{revtex4}
\usepackage{multirow}
\usepackage{dcolumn,graphicx,amsmath,amssymb,algpseudocode,physics}
\usepackage{simplewick}
\usepackage{todonotes}
\usepackage{rotating}
\usepackage{pdflscape}
\usepackage{hyperref}
\usepackage{setspace}
\usepackage{algorithm}
\usepackage{adjustbox}

\begin{document}

\title{Shadow excited state molecular dynamics with the $\Delta$SCF method}

\author{O. Jonathan Fajen}
\affiliation{Department of Chemistry and PULSE Institute, Stanford University, Stanford, CA 94305, United States}
\affiliation{SLAC National Accelerator Laboratory, Menlo Park, CA 94025, United States}
\affiliation{Theoretical Division, Los Alamos National Laboratory, Los Alamos, New Mexico 87545, United States}

\author{Oscar Gr\aa n\"as}
\affiliation{Division of Materials Theory, Department of Physics and Astronomy, Uppsala University, Box 516, SE-75120 Uppsala, Sweden}

\author{Todd J. Mart\'inez}
\affiliation{Department of Chemistry and PULSE Institute, Stanford University, Stanford, CA 94305, United States}
\affiliation{SLAC National Accelerator Laboratory, Menlo Park, CA 94025, United States}

\author{Anders M. N. Niklasson}
\affiliation{Theoretical Division, Los Alamos National Laboratory, Los Alamos, New Mexico 87545, United States}

\begin{abstract}
    We present an extension of the shadow extended Lagrangian Born-Oppenheimer molecular dynamics (XL-BOMD) method to excited state molecular dynamics (ESMD) in the context of $\Delta$SCF Kohn-Sham density functional theory, with demonstrations performed using self-consistent charge density functional tight binding (SCC-DFTB) theory. In this shadow ESMD approach, the approximate iterative solution to the exact potential in conventional ESMD is replaced by an exact single-step solution to an approximate shadow excited-state potential. We show that in addition to offering significant improvement in computational cost relative to direct ESMD, our shadow ESMD method provides enhanced stability and robustness relative to its 'exact' counterpart. Our implementation is carried out in the context of SCC-DFTB theory but should be broadly generalizable, both to {\textit{ab initio}} electronic structure methods and to other semi-empirical quantum chemistry approaches.
\end{abstract}

\maketitle

\section{Introduction}

        The study of molecular dynamics on electronic excited surfaces is of central importance in fields such as photochemistry, \cite{levine_isomerization_2007,AkimovRev} energy storage and conversion, \cite{hammes-schiffer_theory_2010} and surface scattering. \cite{wodtke__electronically_2004,mukherjee_beyond_2019} While many challenges in these areas stem from nonadiabatic effects (involving multiple strongly interacting electronic surfaces), significant insight can be gained from the observation of the dynamics on a single excited state surface, such as the analysis of initial photorelaxation pathways.\cite{frank_molecular_1998,birgisson_decoherence_2025} The ability to accurately describe and follow a potential energy surface is also of fundamental importance to trajectory-based techniques for studying non-adiabatic dynamics, like Tully's surface hopping methods,  \cite{JCTully1971,JCTully12} Ehrenfest dynamics \cite{li_ab_2005}, \textit{ab initio} multiple spawning \cite{ben-nun_ab_2000,prigogine_ab_2002,curchod_ab_2018,crespo-otero_recent_2018}, and its cloning variant \cite{makhov_ab_2014}. While it is possible to carry out excited state molecular dynamics (ESMD) simulations with highly-accurate \textit{ab initio} correlated methods, such as equation-of-motion coupled cluster with single and double excitations (EOM-CCSD), \cite{baeck_ab_2003,choi_ab_2004} \cite{hait_prediction_2024,kjonstad_photoinduced_2024} or complete active space perturbation theory of second order (CASPT2), \cite{coe_ab_2007}\cite{park_caspt2_2017,polyak_ultrafast_2019} the computational cost of these methods becomes prohibitive for systems with more than around 20 heavy atoms. Substantial work has been dedicated to reduce the cost and even the formal scaling of these methods. \cite{hohenstein_tensor_2012,hohenstein_quartic_2013,parrish_tensor_2012,hohenstein_tensor_2013,hohenstein_rank_2019,hohenstein_rank-reduced_2022,song_reduced_2018,fales_large-scale_2018} However, despite these advancements, the less expensive linear-response time-dependent density functional theory (LR-TDDFT), in its random phase approximation (RPA)\cite{casida_chapter,casida_molecular_1998} and Tamm-Dancoff approximation (TDA) variants,\cite{hirata_time-dependent_1999} has become a widely used method in the study of excited state dynamics \cite{TNelson20}. By using various levels of approximation for the exchange-correlation functional in Kohn-Sham density functional theory (KS-DFT), it is sometimes possible to obtain good performance with LR-TDDFT for relevant excitation energies while maintaining computational affordability. However, LR-TDDFT has difficulties with charge-transfer and Rydberg-type excitations and fails to adequately describe ground-excited state conical intersections, preventing its use for simulating full photorelaxation dynamics. To address the latter limitation, methods that treat the ground and excited states on equal footing are required, such as the complete active subspace self-consistent field (CASSCF) method,\cite{roos1980} spin-flip (SF) variants of TDDFT,\cite{krylov_size_consistent_2001,lee_eliminating_2018,lee_efficient_2019}, the particle-particle RPA (pp-RPA),\cite{van_aggelen_exchange-correlation_2013,yang_double_2013} and notably its hole-hole TDA approximation (hh-TDA) \cite{bannwarth_holehole_2020}.
 
        In contrast to LR-TDDFT, excited states can also be targeted by time-independent direct self-consistent field (SCF) optimization, subject to appropriate constraints. This approach, known as $\Delta$SCF, \cite{Bagus65,Slater1972,OGunnarsson76,Gunnarsson89} (or orbital optimization in some contexts), has historically been less explored but has gained significant interest in recent years.\cite{BILundqvist04,Gavnholt08,TZiegler09,TKowalczyk11,RMaurer11, TKowalczyk16,WMackrodt18,Hait2020, DHait21,EVandaele22, TKowalczyk23, JKahk23,WYang24,HPham25} In particular, $\Delta$SCF has been observed to offer better accuracy than LR-TDDFT, especially for charge-transfer-type excitations. \cite{gilbert_self-consistent_2008,hait_orbital_2021} $\Delta$SCF incorporates orbital relaxation into the excited state calculation directly, which may also contribute to its much lower degree of density-functional approximation dependence than TDDFT \cite{DHait21}. Despite these advantages, $\Delta$SCF presents some difficulties in simulation of dynamics occurring on multiple surfaces, as each optimization targets a single state and the resulting states are therefore not orthogonal. The $\Delta$SCF method typically also suffers from convergence problems, such as when excited states are nearly degenerate in energy. This is particularly limiting for molecular dynamics simulations and often leads to stability problems that prevent meaningful simulations.
        These challenges have led to widespread use of 0-th order $\Delta$SCF simulations, where the orthogonal Kohn-Sham orbitals of the ground-state density are used also to represent excited states. \cite{PrezhdoPRL, AkimovRev, pyxaid} 
        When combined with the so-called classical path approximation, that is, assuming that there is no back-reaction from the electronic excited system, efficient surface hopping algorithms may be implemented for complex systems \cite{pyxaid, neukirch, RaquelEP}. However, these simulations rely on the trajectories of an unconstrained ground-state Born-Oppenheimer MD (BOMD).
        
\begin{figure}
            \centering
            \includegraphics[width=0.98\linewidth]{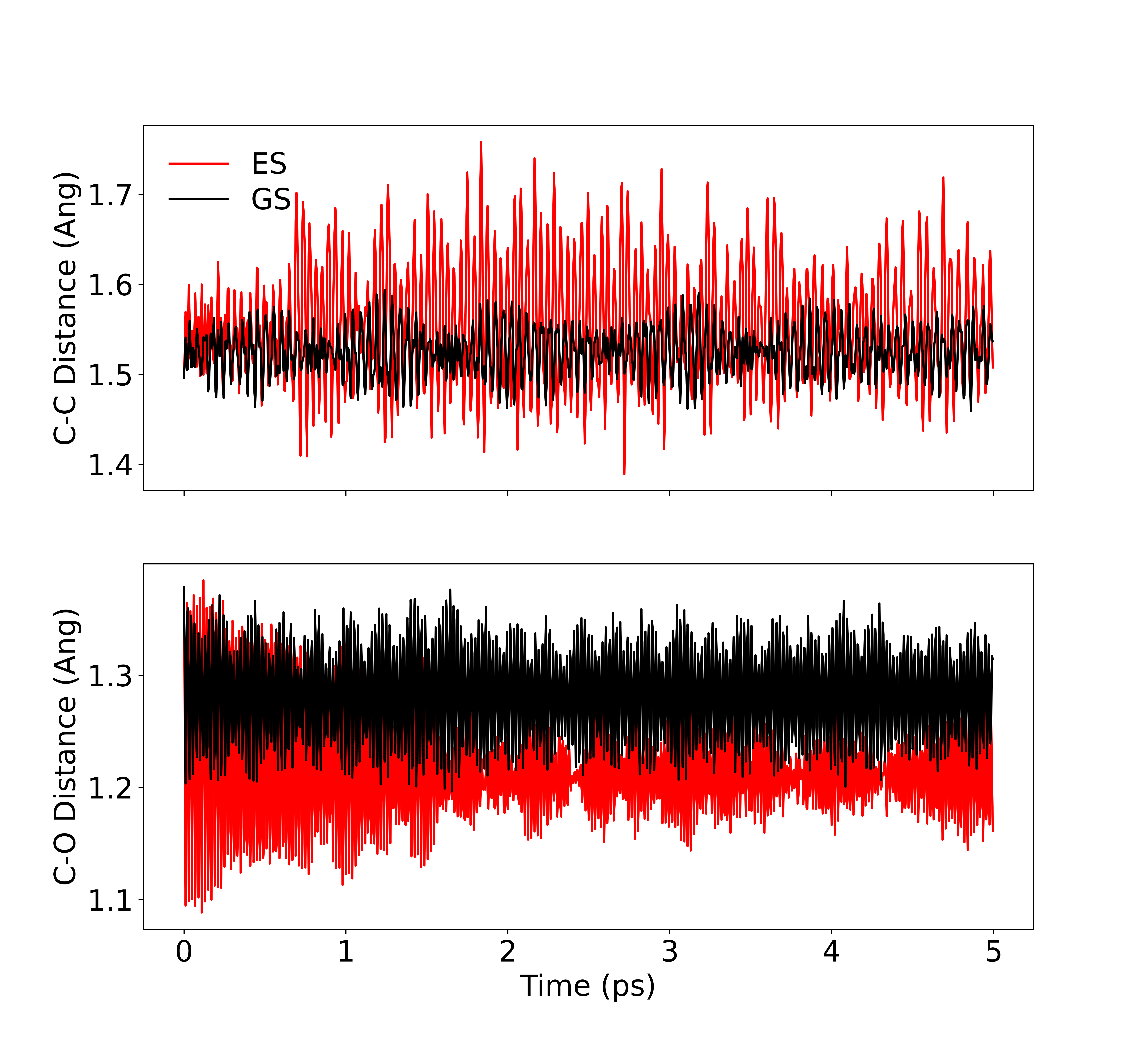}
            \caption{C-C and C-O bond lengths of acetone over a 2 picosecond simulation with SCC-DFTB (details below). The ground state trajectory, labeled GS, is calculated with spin-restricted SCC-DFTB in an XL-BOMD framework. The excited state trajectory, labeled ES, is calculated with an IMOM/quasi-Newton SCC-$\Delta$DFTB scheme, as described later on. The excited state targeted is the lowest singlet corresponding to an excitation from HOMO to LUMO.}
            \label{fig:gs-vs-es}
        \end{figure}

        The development of a robust and computationally efficient approach to follow an excited state potential energy surface described by fully relaxed $\Delta$SCF calculations, while avoiding current instability problems, would be an important advancement. For single-surface dynamics, $\Delta$SCF excited-state molecular dynamics (ESMD) can essentially be viewed as a generalization of ground-state BOMD, with a new non-{\it aufbau} occupation constraint on the electronic density. A self-consistent excited-state density also allows for the full treatment of back-reaction, effectively alleviating issues with the classical path approximation while providing accurate dynamics by following the trajectory generated by the forces of the excited state surface. This is of significant importance as the trajectories of the ground and excited states can be quite different. A relatively mild example of this is illustrated in Fig.\ \ref{fig:gs-vs-es}, which shows the difference between the ground and first singlet excited state trajectories in the oscillations of two bond lengths of acetone over two picoseconds of dynamics. More extreme examples, such as predissociative excited states, only serve to underline the wide discrepancy that typically exists between the potential energy surfaces of different electronic states. Because electronic excitations are typically vertical in character, the nuclei experience very different forces on the ground and excited state potential energy surfaces. For example, as seen in Fig.\ \ref{fig:gs-vs-es}, the excited state trajectory displays larger oscillations in the two bond lengths in the acetone simulation compared to the ground state. In general, the dynamics of the nuclei will respond drastically after undergoing near instantaneous promotion to an electronic excited surface.

        The computational cost of ground-state BOMD is dominated by the iterative self-consistent optimization of the electronic density at each timestep, which is required prior to the force evaluations. This process can be greatly accelerated by a judicious choice of initial guess. It is therefore unsurprising that much work has been done to improve the quality of initial guesses in ground state BOMD simulations \cite{MCPayne92,PPulay04,JMHerbert05,ANiklasson06,TDKuhne07,JFang16,EPolack21}. 
        
        Reusing the converged density of the previous timestep is the most straightforward way of obtaining a good initial guess in BOMD and has long been known to accelerate convergence. However, if very tight SCF convergence thresholds are not enforced, simply reusing the old density or extrapolating previous densities as an initial guess will introduce a bias to the electronic degrees of freedom, leading to a systematic energy drift over time \cite{DRemler90,PPulay04}. It is possible to avoid this systematic drift by propagating a time-reversible linear combination of previous initial guesses and converged densities \cite{ANiklasson06}, but even better performance- both in terms of computational efficiency and simulation stability- can be obtained by propagating the electronic density as an additional dynamical field variable. This approach is used in extended Lagrangian Car-Parrinello molecular dynamics (CPMD) and extended Lagrangian BOMD (XL-BOMD) \cite{RCar85,JHutter12,ANiklasson08}.
 
        In its modern formulation, XL-BOMD emphasizes that the stability and physical consistency between the forces and the potential energy surface are of key importance to accurate dynamics \cite{ANiklasson17,ANiklasson20,ANiklasson21b,Niklasson23}. By constructing an approximate 'shadow' potential energy surface on which the electronic degrees of freedom can be propagated exactly (in an adiabatic limit), the need for iterative SCF optimization can be eliminated. This dramatically lowers the cost of ground-state BOMD simulations while at the same time improving the numerical stability of the propagation. 
        
        The concept of a shadow dynamics, or a shadow Hamiltonian, is well established in classical dynamics. It has been used to analyze and construct new integration schemes that often exhibit superior properties \cite{HYoshida90,CGrebogi90,SToxvaerd94,JGans00,BJLeimkuhler94,RDEngel05,BJLeimkuhler04,ShadowHamiltonian,SToxvaerd12,KDHammonds21}. The shadow molecular dynamics approach has also shown significant success in ground-state XL-BOMD simulations, even for highly unstable reactive chemical systems \cite{ANiklasson20,ANiklasson21b,CNegre22,Niklasson23}. 
        As an extension of BOMD to the excited state, a $\Delta$SCF-based shadow ESMD would be expected to inherit the benefits of XL-BOMD in the ground state, i.e.\ greatly reduced computational cost and enhanced stability. The shadow XL-BOMD approach could be even more appealing in the ESMD context, as the SCF problem for excited states can be more challenging than for the ground state.  
        
        The main goal of this paper is to introduce a $\Delta$SCF-based shadow ESMD within an extended Lagrangian framework and to explore if this is a viable path forward for MD simulations on excited state potential energy surfaces. 
        
        The XL approach has previously been applied to ESMD using the RPA equations \cite{JBjorgaard18}. In that work, the transition density matrix was added to the extended Lagrangian as an additional dynamical variable and substantial computational savings were observed, as well as the elimination of any systematic energy drift. However, this method did not constitute a shadow dynamics and a few Davidson iterations were still required each timestep. In the present work, we introduce a more rigorous shadow ESMD approach based on $\Delta$SCF Kohn-Sham DFT for efficient excited state molecular dynamics simulations within an XL framework. Our proposed XL-$\Delta$SCF-based shadow ESMD is demonstrated in the context of self-consistent charge density functional tight-binding (SCC-DFTB) theory \cite{MElstner98,MFinnis98,TFrauenheim00} and we explore its potential as an accelerated ESMD method. The XL shadow MD framework circumvents many of the stability issues of the SCF cycle common to the $\Delta$SCF procedure for complex systems. Furthermore, near-degeneracies in the Kohn-Sham Hamiltonian do not pose a severe issue, because the potential energy surface of the extended Lagrangian traces the character of the initially converged $\Delta$SCF state in the first time step. This ensures a more stable and consistent propagation throughout the simulation.

\section{Theory}

    \subsection{Spin-polarized KS-DFT BOMD}

    We begin by describing the relevant parts of ground-state BOMD for spin-polarized KS-DFT. The ground-state Kohn-Sham spin densities \cite{hohen,KohnSham65,Barth_1972,Gunnarsson89,RMDreizler90} are given by a sum over the partial densities from the molecular spin orbitals,
    \begin{equation}
        \rho^{\sigma}({\bf r}) = \sum_{i} f_{i}^{\sigma} | \psi_{i}^{\sigma}(\textbf{r}) |^2.
    \end{equation}
    Here $f_{i}^{\sigma}$ are the orbital occupation numbers for each spin channel, $\sigma = \{ \alpha, \beta \}$ and the $\psi_{i}^{\sigma}(\textbf{r})$ are the normalized Kohn-Sham spin orbitals, i.e.\ where
    \begin{equation}
        \int \textit{d} \textbf{r }  |\psi_{i}^{\sigma}(\textbf{r}) |^2 = 1,~ \textbf{ }\forall \textit{ i }.
    \end{equation}
    The ground state Born-Oppenheimer (BO) potential energy function is then given by the constrained minimization,
    \begin{equation}
        U_{\rm BO}(\textbf{R}) = \min_{\rho^{\sigma},f^{\sigma}} \Big \{ F_{\rm KS}[ \rho^{\sigma} ] + \sum_{\sigma} \int d\textbf{r } V_{\rm ext}^{\sigma} (\textbf{R},\textbf{r}) \rho^{\sigma}(\textbf{r}) ~\Big | ~\sum_{i,\sigma} f_{i}^{\sigma} = N_{e} \Big \} + V(\textbf{R}). \label{BO_pot_opt}
    \end{equation}
    The occupation numbers, $f_i^\sigma$, are 0 for the unoccupied states and 1 for the occupied. To minimize the energy, the occupation follows the {\it aufbau} principle occupying the states from the lowest to the highest energies until the sum is the total number of electrons, $N_e$. In this article we will not consider fractional occupation numbers, which also would require an additional electronic entropy term to account for the free energy \cite{RWentzcovitch92,RParr89}. $V(\textbf{R})$ is the nuclear repulsion energy for the atoms at positions ${\bf R} = \{{\bf R}_I\}$. $V_{\rm ext}^{\sigma}({\bf R,r})$ is a spin-dependent external potential and $F_{\rm KS}[\rho^
    \sigma]$ is the Kohn-Sham spin-density energy functional. The dynamics of this Born-Oppenheimer potential can be defined by the Lagrangian,
    \begin{equation}
        L_{\rm BO}(\textbf{R},\dot{\textbf{R}}) = \frac{1}{2} \sum_{I} m_{I} |\dot{\textbf{R}}_{I} | ^{2} - U_{\rm BO}(\textbf{R}),
    \end{equation}
    with the corresponding Euler-Lagrange's equations of motion,
    \begin{equation}
        m_{I} \Ddot{\textbf{R}}_{I} = -\nabla_{I} U_{\rm BO}(\textbf{R}). \label{Newton}
    \end{equation}
    Here $m_I$ are the atomic masses and the dots denote the time derivatives.
    To generate the molecular trajectories, the equations of motion can be integrated using, for example, the leapfrog velocity Verlet algorithm. 
    
    The main cost of this BOMD simulation scheme is the constrained minimization in Eq.\ (\ref{BO_pot_opt}). This is a non-linear minimization problem that has to be solved iteratively until a sufficiently converged self-consistent solution is found for the relaxed electronic ground state. Unless the solution is well converged, the forces in Eq.\ (\ref{Newton}) may not be sufficiently accurate and conservative. This is a particularly difficult problem when the ground state minimization in Eq.\ (\ref{BO_pot_opt}) is accelerated by using a good initial guess extrapolated from previous timesteps during an MD simulation. In this case insufficiently converged ground state solutions will lead to a systematic energy drift, where the electronic solution behaves like an artificial heat-sink or heat-bath \cite{DRemler90,PPulay04}. Performing a BOMD simulation in a canonical ensemble, using a thermostat, does not avoid the underlying problem \cite{EMartinez15}.

    \subsection{Excited state MD with $\Delta$SCF}

    \begin{figure}
        \centering
        \includegraphics[width=0.5\linewidth]{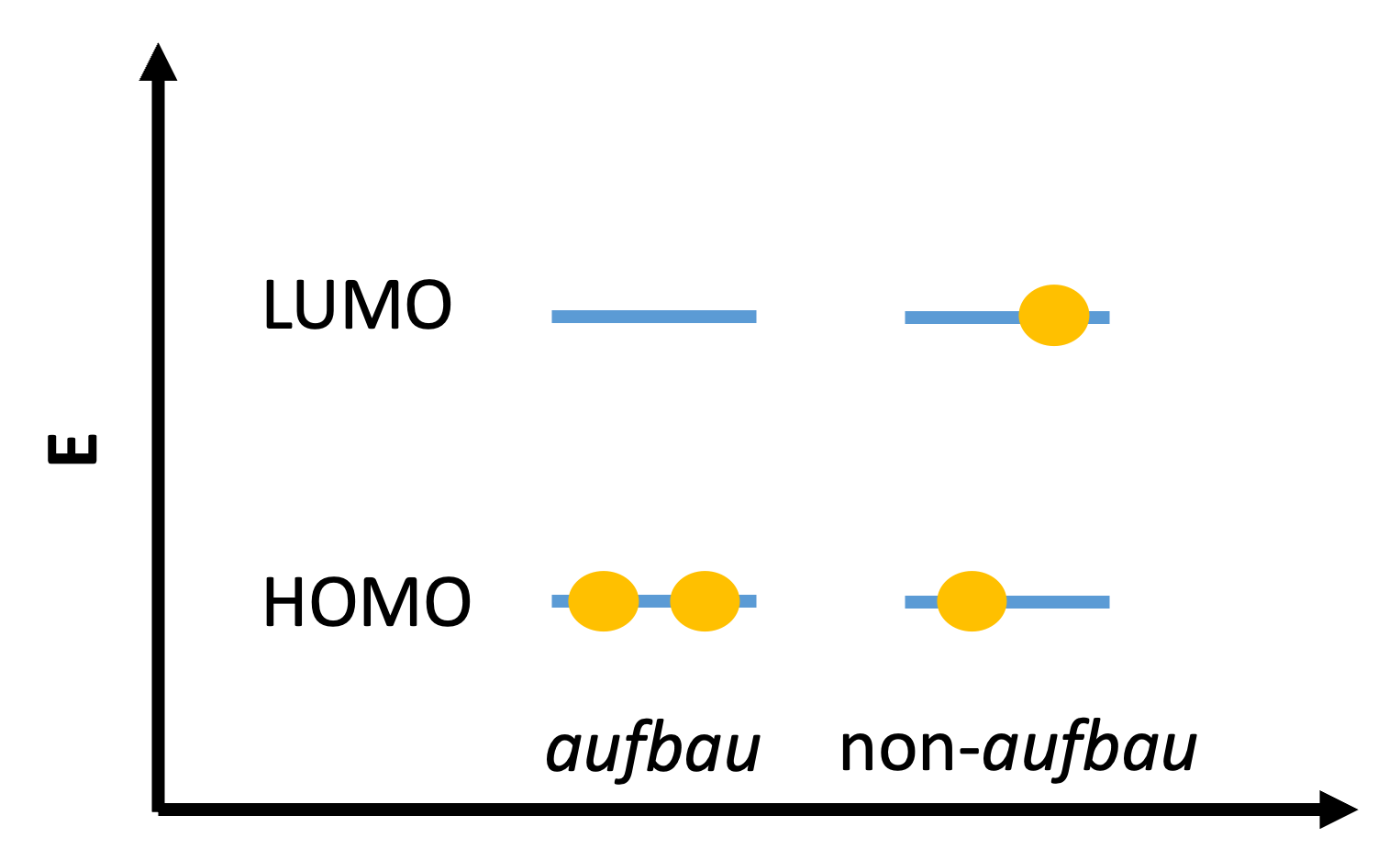}
        \caption{Pictorial representation of the \textit{aufbau} and non-\textit{aufbau} configurations in the basis of highest occupied and lowest unoccupied molecular orbitals (HOMO and LUMO) of the ground state, corresponding to an alpha occupation list: $\rm{occ}(\alpha): (1,2,...,HOMO-1,LUMO)$, and beta occupation list: $\rm{occ}(\beta): (1,2,...,HOMO-1,HOMO)$. Higher-lying excitations or double excitations can also be generated in this way.}
        \label{fig:nonaufbau}
    \end{figure}

    Moving from ground state KS-DFT to excited state $\Delta$SCF KS-DFT requires changing the {\it aufbau} occupation constraint for the ground state to a non-{\it aufbau} occupation constraint for the excited state, as depicted in Fig.\ \ref{fig:nonaufbau}. In this way we can define a generalized $\Delta$SCF potential energy surface,
    \begin{equation}
        U_{\rm \Delta SCF}({\bf R}) = \min_{\rho^{\sigma}} \Big \{ F_{\rm KS} [\rho^{\sigma}] + \sum_{\sigma} \int d\textbf{r }V_{\rm ext}^{\sigma} \rho^{\sigma} (\textbf{r}) ~\Big |~ {\bf f}^{\sigma} \in \Omega_{\sigma}  \Big \} + V(\textbf{R}), \label{U_DeltaSCF}
    \end{equation}
    with the corresponding optimized excited state density, $\rho_{\rm min}^{\sigma} (\textbf{r})$.
    Here ${\bf f}^\sigma \in \Omega_{\sigma}$ represent the non-{\it aufbau} constraints of the occupation numbers, ${\bf f}^\sigma \equiv \{f^\sigma_i\}$, imposing a non-{\it aufbau} occupation of the KS spin orbitals along with ensuring that the occupation numbers sum to $N_{e}$. The {\it aufbau} vs. non-{\it aufbau} occupation constraints are illustrated in Fig.\ \ref{fig:nonaufbau} for the highest occupied and lowest unoccupied molecular orbitals (HOMO and LUMO) of the relaxed ground state. In this work we generally target the lowest singlet excited state with $\Delta$SCF by enforcing the occupation of the ground state HOMO to be 0 and that of the ground state LUMO to be 1 in the alpha spin channel (and vice-versa for the beta spin channel). 

    If we view $\Delta$SCF from the perspective of a single-determinant wavefunction theory, this approach will introduce spin-contamination of singlet excited states. For instance, the configuration HOMO-1($\alpha$, $\beta$), HOMO($\beta$), LUMO($\alpha$) is not a pure spin eigenstate, but rather a combination of singlet and $M_{S} = 0$ triplet. In a wavefunction theory, the Ziegler sum correction would exactly ameliorate this problem at the cost of an additional calculation of the triplet energy, and this correction has previously been used with success in the context of DFT \cite{ziegler_implementation_2012,bourne_worster_reliable_2021,TKowalczyk11} and DFTB. \cite{zheng_implementation_2009,TKowalczyk16} In our demonstration of our shadow $\Delta$SCF ESMD we will not pursue the Ziegler sum correction, although its implementation would be straightforward.

The $\Delta$SCF optimization in Eq.\ (\ref{U_DeltaSCF}) begins by asserting a non-{\it aufbau} occupation of the self-consistent ground state Kohn-Sham orbitals. In the most straightfoward algorithm for the $\Delta$SCF optimization, the occupation numbers are simply adjusted each SCF cycle to maintain the same configuration. This approach can run into difficulties such as when the KS orbital ordering changes between cycles, leading to variational collapse to the ground state or other convergence failures. These problems can be ameliorated to some extent by the Maximum Overlap Method (MOM) \cite{gilbert_self-consistent_2008} or Initial Maximum Overlap Method (IMOM) approaches of Gill and co-workers, or more recent methods, \cite{carter-fenk_state-targeted_2020,ivanov_method_2021,schmerwitz_variational_2022,schmerwitz_calculations_2023,schmerwitz_saddle_2024,sigurdarson_orbital-optimized_2023} notably including Squared Gradient Minimization (SGM). \cite{hait_excited_2020,hait_orbital_2021} or direct energy minimization techniques \cite{GLevi20,HPham25}. The MOM differs from the naive energy-based occupation reordering by instead computing the overlap of the new and old KS orbitals,
\begin{equation}
    O = (C^{\rm old})^{\dagger}SC^{\rm new}.
\end{equation}
The $N_{e}$ orbitals with greatest projection onto the previously occupied orbitals, $p_{j} = \sum_{i} (O_{ij}^{2})^{\frac{1}{2}}$, are occupied each cycle. The IMOM procedure is identical except that $C^{\rm old}$ is replaced by $C^{\rm initial}$ in every cycle, where $C^{\rm initial}$ is the new initial SCF guess for each new MD time step. This can avoid drift away from the starting configuration, which is assumed to be a good representation of the non-{\it aufbau} configuration, although convergence failures can still occur. In the present work, we will employ the IMOM scheme in some of our reference ESMD simulations.

The dynamics for the $\Delta$SCF potential in Eq.\ (\ref{U_DeltaSCF}) can then be described by the Lagrangian,
    \begin{equation}
        L_{\rm \Delta SCF} (\textbf{R},\dot{\textbf{R}}) = \frac{1}{2} \sum_{I} m_{I} | \dot{\textbf{R}}_{I} |^2 - U_{\rm \Delta SCF}(\textbf{R}),
    \end{equation}
with the Euler-Lagrange's equations of motion,
    \begin{equation}
           m_I  \Ddot{\textbf{R}}_{I} = -\nabla_{I} U_{\rm \Delta SCF} (\textbf{R}).
    \end{equation}
As for the previous ground-state Born-Oppenheimer molecular dynamics scheme, the main cost of $\Delta$SCF-based molecular dynamics simulations is the constrained non-linear minimization in Eq.\ (\ref{U_DeltaSCF}), which is required in each integration time step. However, in contrast to the ground state ({\it aufbau}) optimization, the iterative $\Delta$SCF optimization for excited states is often notably harder, including the potential for variational collapse to the ground state. The convergence and instability problems for ground state Born-Oppenheimer MD simulations are therefore more pronounced for the excited state. The main goal of this article is to introduce a shadow ESMD based on $\Delta$SCF within an extended Lagrangian framework and investigate its ability to reduce the computational hurdles and shortcomings for excited state $\Delta$SCF MD simulations in the same way as has been possible for ground state shadow XL-BOMD. 

    \subsection{Shadow ESMD with $\Delta$SCF}

    Our shadow molecular dynamics is based on an approximate shadow potential. The idea behind this concept is simple, but quite powerful. Instead of constructing the potential from an approximately optimized exact energy functional, we define an approximate \textit{shadow} potential from an exactly optimized but approximate \textit{shadow} energy functional. For the $\Delta$SCF shadow ESMD we accomplish this by linearizing the energy functional, $F_{\rm KS} \rightarrow \mathcal{F}_{\rm KS}$, about an approximate excited state density, $n^{\sigma}(\textbf{r}) \approx \rho_{\rm min}^{\sigma}(\textbf{r})$, where
    \begin{equation}
        \mathcal{F}_{\rm KS}[\rho^{\sigma},n^{\sigma}] = F_{\rm KS} [n^{\sigma}] + \sum_{\sigma} \int d\textbf{r } (\rho^{\sigma} (\textbf{r})-n^{\sigma}(\textbf{r})) \frac{\delta F_{KS}[\{\rho^{\alpha},\rho^{\beta} \}]}{\delta \rho^{\sigma}(\textbf{r})} \Big |_{\rho^{\sigma}=n^{\sigma}}.
    \end{equation}
 Our shadow excited state potential, $\mathcal{U}_{\rm \Delta SCF} (\textbf{R},n^{\sigma} )$, is then $n^{\sigma}$-dependent and is determined by the lowest stationary solution to the constrained minimization (or equilibration),
    \begin{equation}
        \mathcal{U}_{\rm \Delta SCF} (\textbf{R},n^{\sigma} ) = \min_{\rho^{\sigma}} \Big \{ \mathcal{F}_{\rm KS} [\rho^{\sigma}, n^{\sigma}] + \sum_{\sigma} \int d\textbf{r }V_{ext}^{\sigma} (\textbf{R},\textbf{r}) \rho^{\sigma} (\textbf{r}) ~\Big | ~ {\bf f}^{\sigma} \in \Omega_{\sigma} \Big \} + V(\textbf{R}),
    \end{equation}
    which is attained at the $n^\sigma$-dependent excited state density, $\rho_{\rm min}^{\sigma} [n^{\sigma}](\textbf{r})$. Because $\mathcal{F}_{\rm KS}[\rho^{\sigma},n^{\sigma}]$ is a linearized functional, this constrained equilibration can be done exactly in a single step, e.g.\ by diagonalizing a Kohn-Sham Hamiltonian that depends on a fixed density, $n^\sigma$, avoiding costly iterations and the associated convergence problems. The correct conservative forces from the gradients of the approximate shadow potential can then be calculated without problems. A shadow ESMD can then be constructed based on $\mathcal{U}_{\rm \Delta SCF} (\textbf{R},n^{\sigma} )$.
    The accuracy of this shadow ESMD depends on the how close $n^{\sigma}({\bf r})$ is to the exact  excited state density, $\rho_{\rm min}^{\sigma} (\textbf{r})$. However, as the atoms are moving during a molecular dynamics simulation the approximate excited state density, $n^{\sigma}(\textbf{r})$, around which we perform the linearization, may no longer be close to the exact solution, $\rho_{\rm min}^{\sigma}(\textbf{r})$, of the regular $\Delta$SCF method. We therefore need to update the approximate excited-state density as the atoms are moving. We can achieve this through an extended Lagrangian formulation of the dynamics, where $n^{\sigma}(\textbf{r})$ is included as a dynamical field variable that is propagated through a harmonic oscillator centered around $\rho_{\rm min}^{\sigma}(\textbf{r})$, or its best available approximation, $\rho_{\rm min}^{\sigma} [n^{\sigma}](\textbf{r})$.
    The corresponding extended Lagrangian for our $\Delta$SCF-based ESMD is defined as
    \begin{equation}
    \begin{split}
        \mathcal{L}_{\rm \Delta SCF} (\textbf{R},\dot{\textbf{R}},n^{\sigma},\dot{n}^{\sigma}) = \frac{1}{2} \sum_{I} m_{I} | \dot{\textbf{R}}_{I} |^{2} - \mathcal{U}_{\rm \Delta SCF}( \textbf{R}, n^{\sigma}) + \frac{1}{2}\mu \sum_{\sigma} \int d\textbf{r } (\dot{n}^{\sigma}(\textbf{r}))^{2} \\ - \frac{1}{2} \mu \omega^{2} \sum_{\sigma,\sigma{'}} \iint d\textbf{r }d\textbf{r}{'}\textbf{ } (\rho_{\rm min}^{\sigma}[n](\textbf{r}) - n^{\sigma}(\textbf{r}) ) T^{\sigma \sigma{'}} (\textbf{r},\textbf{r}{'}) (\rho_{\rm min}^{\sigma{'}}[n](\textbf{r}{'}) - n^{\sigma{'}}(\textbf{r}{'})).
    \end{split}
    \end{equation}
    The extended $\Delta$SCF Lagrangian includes the electronic density, $n^{\sigma}$, as an additional dynamical variable, together with an extended harmonic oscillator, which keeps $n^{\sigma}$ oscillating close to the optimal excited state density, $\rho_{\rm min}^{\sigma} [n^{\sigma}](\textbf{r})$. Here $\mu$ is a fictitious electronic mass and $\omega$ is the frequency of the extended harmonic oscillator. The metric tensor of the harmonic oscillator, $T^{\sigma \sigma{'}} (\textbf{r},\textbf{r}{'})$, is symmetric positive definite and is determined by a kernel function,

    \begin{equation}
        T^{\sigma \sigma{'}} = \sum_{\sigma{''}} \int d\textbf{r}{''}\textbf{ } K^{\sigma{''} \sigma} (\textbf{r}{''},\textbf{r}) K^{\sigma{''} \sigma{'}} (\textbf{r}{''}\textbf{r}{'}).
    \end{equation}
    The kernel, $K^{\sigma{'} \sigma} (\textbf{r}{'},\textbf{r})$, is taken as the inverse of the Jacobian, $J^{\sigma{'} \sigma}({\bf r},{\bf r}')$, of the spin-density (charge) residual function,
    \begin{equation}
    J^{\sigma \sigma{'}}({\bf r},{\bf r}')  = \frac{\delta \left(\rho_{\rm min}^{\sigma}[n](\textbf{r}) - n^{\sigma}(\textbf{r}) \right)}{\delta n^{\sigma'}({\bf r}')}.
    \end{equation}
    Our extended Lagrangian then leads to a dynamics determined by the Euler-Lagrange equations of motion for the nuclei, as well as the dynamical electronic density. The equations of motion are determined in a classical adiabatic limit, where we assume that the extended electronic degrees of freedom are fast compared to the slower nuclear motion. In this limit, we assume that $\vert q_{{\rm min},J}^{\sigma{'}}[n]-n_{J}^{\sigma{'}} \vert \propto \omega^{-2}$, and we take the limit $\mu \rightarrow 0$ such that the product $\mu \omega$ stays constant \cite{ANiklasson21b}. In this zero-mass Born-Oppenheimer-like approximation we get the coupled equations of motion,
    \begin{equation}
        m_{I} \ddot{\textbf{R}}_{I} = -\nabla_{I} \mathcal{U}_{\rm \Delta SCF} (\textbf{R},n^{\sigma}) \Big |_{n^{\sigma}},
    \end{equation}

    \begin{equation}
        \ddot{n}^{\sigma}(\textbf{r}) = -\omega^{2} \sum_{\sigma{'}} \int d\textbf{r }K^{\sigma \sigma{'}} (\textbf{r},\textbf{r}{'}) \left(\rho_{\rm min}^{\sigma{'}}[n](\textbf{r}) -n^{\sigma{'}}(\textbf{r})\right),
    \end{equation}
    which are exact in continuous time. The first equation describes the motion of the nuclei on the excited-state shadow potential energy surface. The gradient is evaluated at constant $n^\sigma$, since $n^\sigma$ is treated as a dynamical field variable. This formulation results in a force expression with a level of simplicity comparable to Hellmann–Feynman forces, while avoiding the need for iterative, nonlinear optimization of the excited-state density at each step. The equation can be integrated using standard schemes, such as the leapfrog velocity Verlet algorithm. The efficiency of the shadow excited-state method comes from the extended Lagrangian dynamics. In a static setting, evaluating the gradient of the shadow potential would require additional terms of the form $(\delta \mathcal{U}_{\Delta {\rm SCF}}/\delta n)(\partial n/\partial \mathbf{R}_i)$, which are computationally expensive. By contrast, the extended Lagrangian approach eliminates this overhead by propagating $n^\sigma({\bf r})$ as a dynamical field variable. The second equation is an equation of motion for the extended harmonic oscillator that drives the evolution of the approximate electron density, $n^\sigma({\bf r})$, around which we linearize the energy functional for the construction of the shadow potential. This propagated density oscillates in a harmonic well that is centered closely around the exact ground state density as the atoms are moving.\ \cite{ANiklasson21b} This equation can be integrated with a regular Verlet integration scheme. However, in order to make sure that the propagated electron density stays synchronized with the evolution of the atomic coordinates, we include an additional weak dissipative force term that also removes accumulation of numerical noise. This modified Verlet integration scheme is described in Refs.\  \cite{ANiklasson09,PSteneteg10,GZheng11,ANiklasson21b}.

Ground state extended Lagrangian Born-Oppenheimer shadow MD simulations circumvents the SCF stability problems even for challenging reactive chemical systems. The main purpose of this article is to extend this approach to $\Delta$SCF-based ESMD simulations, where the stability problems often are prohibitively difficult to overcome. To evaluate if this new shadow ESMD with the $\Delta$SCF method can alleviate the stability problems we have implemented the method in combination with self-consistent-charge density functional tight-binding (SCC-DFTB) theory \cite{MFoulkes89,DPorezag95, MElstner98,MFinnis98,TFrauenheim00,BAradi07, BHourahine20}.  

    \subsection{SCC-DFTB}

    As a test model to evaluate and demonstrate the $\Delta$SCF-based shadow ESMD, we have implemented the method using SCC-DFTB theory for the electronic structure \cite{MFoulkes89,DPorezag95, MElstner98,MFinnis98,TFrauenheim00,BAradi07, BHourahine20}. SCC-DFTB theory is a fairly inexpensive approximation to KS-DFT and can be readily parameterized to improve performance for particular properties or subsets of chemical space. This is particularly appealing in combination with modern machine learning methods \cite{PDral2015, DYaron18,PDral20,ZGuoqing22,DYaron23}. To combine the $\Delta$SCF method with SCC-DFTB theory is not new and the performance of this $\Delta$DFTB theory for excitation energies and excited state properties has been studied in the past  \cite{TKowalczyk16,TKowalczyk23}.  Our aim in this work in not to present an optimal tuning of SCC-DFTB parameters to achieve high-level excited state accuracy, or to demonstrate the capability for large-scale, high-throughput simulations, but rather to analyze and demonstrate the potential gains provided by our shadow ESMD approach over regular (direct) ESMD. 

    Our implementation is based on the DFTB code LATTE \cite{LATTE,MCawkwell12,AKrishnapriyan17}. The ground state spin-polarized SCC-DFTB Born-Oppenheimer potential is given by  
    \begin{equation}
        U_{\rm BO}^{\rm DFTB} ( \textbf{R} ) = \min_{D^{\sigma},f^{\sigma}} \Big \{ \sum_{\sigma} {\rm Tr}[ H_{\sigma}^{(0)}(D^{\sigma}-D_{0}^{\sigma}) ] + \frac{1}{2} \sum_{I,J} q_{I} \gamma_{IJ} q_{J} + \frac{1}{2} \sum_{I} M_{I}(q_{I}^{\alpha} - q_{I}^{\beta})^{2} ~\Big | ~\sum_{i,\sigma} f_{i}^{\sigma} = N_{e} \Big \} + V_{\rm ref} (\textbf{R})
    \end{equation}
    Here we use the atom-resolved, onsite spin interaction parameters, $M_I$, of Melix and co-workers. \cite{melix_spin_2016} A more sophisticated approach might incorporate orbital-resolved spin parameters, higher-order multipoles \cite{VQVuong23}, non-local dispersion interactions \cite{BHourahine20}, and more advanced many-body expansions \cite{RKLindsey17,NGoldman23} of the repulsive reference potential, $V_{\rm ref}({\bf R})$, but we have opted for a simpler coarse treatment in this initial study. The net partial charge of each atom is determined by the Mulliken population as given from the net spin-densities, where  
\begin{equation}
    q_{I} = \sum_{\sigma} {\rm Tr}[(D^{\sigma}-D_{0}^{\sigma})S_{I}].
\end{equation}
Here $D^{\sigma}$ is the spin-density matrix, $D_{0}^{\sigma}$ is the neutral, non-interacting atomic (spin) density matrix, and $S_I$ is the symmetrized partial overlap matrix between atomic site $I$ and all the other atoms $J$, i.e. 
\begin{equation}
    S_{I}{'} = \{ S_{ij}\}_{i\in I,j\in J} \textit{, } S_{I} = \frac{1}{2} (S_{I}{'} + {S_{I}{'}}^{T}).
\end{equation}
The matrix $\gamma_{IJ}$ represents a screened Coulomb interaction between overlapping atom-centered charge distributions \cite{MElstner98}. When $I = J$, the Coulomb interaction takes the form of a Hubbard-U onsite Coulomb repulsion term and decays like $|{R_{I}-R_{J}}|^{-1}$ at larger atomic separations, i.e.
\begin{equation}
    \gamma_{IJ} = 
    \begin{cases}
        |R_{I}-R_{J}|^{-1} & |R_{I}-R_{J}| \rightarrow \infty\\
        U_{IJ} & |R_{I}-R_{J}| \rightarrow 0.
    \end{cases}
\end{equation}
The spin-dependent effective Hamiltonian takes the form,

\begin{equation}
    H_{\sigma} = H^{(0)} + \frac{1}{2} (V^{C}S + S V^{C}) + \frac{1}{2} (W^{\sigma}S + SW^{\sigma}),
\end{equation}
where $H^{(0)}$ is a charge-independent Slater-Koster tight-binding Hamiltonian, and $V^{C}$ and $W^{\sigma}$ are effective Coulomb and spin interaction matrices defined below. The orthogonal basis representation of the Hamiltonian is obtained by

\begin{equation}
    H_{\sigma}^{\perp} = Z^{T}H_{\sigma}Z,  
\end{equation}
with the congruence transformation determined by
\begin{equation}
    Z^{T}SZ = I.
\end{equation}
The orthogonal basis Hamiltonian has eigenvalues and eigenvectors determined by the eigenvalue equation,
\begin{equation}
    H_{\sigma}^{\perp} c_{i}^{\sigma} = \epsilon_{i}^{\sigma} c_{i}^{\sigma},
\end{equation}
and the corresponding density matrices are given by
\begin{equation}
    D^{\sigma} = Z (\sum_{i} f_{i}^{\sigma} c_{i}^{\sigma} {c_{i}^{\sigma}}^{T}) Z^{T}.
\end{equation}
The effective Coulomb interaction matrix is given by
\begin{equation}
    V_{i\in I,j\in J}^{C} = v_{I}^{C} \delta_{IJ}, ~{\rm where~} v_{I}^{C} = \sum_{J} \gamma_{IJ} q_{J},
\end{equation}
and the spin interaction matrices are given in the $\alpha$ case by
\begin{equation}
    W_{i\in I,j \in J}^{\alpha} = w_{I}^{\alpha} \delta_{IJ}, ~{\rm where~}
    w_{I}^{\alpha} = M_{I}(q_{I}^{\alpha}-q_{I}^{\beta})
\end{equation}
Analogously for $\beta$ we have
\begin{equation}
    W_{i\in I,j \in J}^{\beta} = w_{I}^{\beta} \delta_{IJ}~{\rm where~}w_{I}^{\beta} = -M_{I}(q_{I}^{\alpha}-q_{I}^{\beta}).
\end{equation}

\subsection{Shadow Excited-state MD with SCC-$\Delta$DFTB}

To obtain a shadow SCC-$\Delta$DFTB potential, we linearize the DFTB energy function about the extended dynamical variables, $n^{\sigma}$, which are assumed to be close to the exact excited state spin charges, yielding
\begin{equation}
\begin{split}
    \mathcal{U}_{\Delta \rm SCF}^{\rm DFTB}(\textbf{R},n^{\sigma}) = \min_{D^{\sigma}} \Big \{  \sum_{\sigma} {\rm Tr}[H_{\sigma}^{(0)}(D^{\sigma}-D_{0}^{\sigma})] + \frac{1}{2} \sum_{I,J} (2q_{I}-n_{I}) \gamma_{IJ} n_{J} \\+ \frac{1}{2} (2(q_{I}^{\alpha}-q_{I}^{\beta})-(n_{I}^{\alpha}-n_{I}^{\beta})) M_{I} (n_{I}^{\alpha}-n_{I}^{\beta}) \big \vert ~ {\bf f}^{\sigma} \in \Omega_{\sigma} \Big \} + V_{\rm ref}\rm (\textbf{R}),
\end{split}
\end{equation}
where the minimum is attained for $n$-dependent charges, $q_{\rm min}^\sigma[n]$.
The attendant extended Lagrangian for SCC-$\Delta$DFTB is then given by

    \begin{equation}
    \begin{split}
        \mathcal{L}_{\Delta \rm SCF}^{\rm DFTB} (\textbf{R},\dot{\textbf{R}},n^{\sigma},\dot{n}^{\sigma}) = \frac{1}{2} \sum_{I} m_{I} | \dot{\textbf{R}}_{I} |^2 - \mathcal{U}_{\Delta \rm SCF}^{\rm DFTB} (\textbf{R},n^{\sigma}) + \frac{1}{2}\mu \sum_{I,\sigma} | \dot{n}_{I}^{\sigma} |^2 \\ - \frac{1}{2} \mu \omega^2 \sum_{I,J,\sigma,\sigma{'}} (q_{{\rm min},I}^{\sigma}[n]-n_{I}^{\sigma})T_{IJ}^{\sigma \sigma{'}} (q_{{\rm min},J}^{\sigma{'}}[n]-n_{J}^{\sigma{'}}). \label{XLDSCF}
        \end{split}
    \end{equation}
Here, the metric tensor of the harmonic potential is $T = K^{T}K$, where the kernel $K$ is given by the inverse of the Jacobian charge residual function, i.e. $K = J^{-1}$, where
\begin{equation}
    J_{IJ}^{\sigma \sigma{'}} = \frac{\partial}{\partial n_{J}^{\sigma{'}}} (q_{{\rm min},I}^{\sigma}[n]-n_{I}^{\sigma}).
\end{equation}

The Euler-Lagrange equations of motion for the Lagrangian in Eq.\ (\ref{XLDSCF}) in the adiabatic limit, where $\omega \rightarrow \infty$, $\mu \rightarrow 0$, as $\mu \omega = {\rm constant}$, and assuming $\vert q_{{\rm min},J}^{\sigma{'}}[n]-n_{J}^{\sigma{'}} \vert \propto \omega^{-2}$, \cite{ANiklasson21b} leads to the equations of motion for our DFTB-based shadow excited state molecular dynamics with the $\Delta$SCF method,
\begin{equation}
    m_{I} \Ddot{\textbf{R}}_{I} = -\nabla_{I} \mathcal{U}_{\Delta \rm SCF}^{\rm DFTB} (\textbf{R},n^{\sigma})|_{n^{\sigma}},
\end{equation}
\begin{equation}
    \Ddot{n}_{I}^{\sigma} = -\omega^{2} \sum_{J,\sigma{'}} K_{IJ}^{\sigma \sigma{'}} \left(q_{{\rm min},J}^{\sigma{'}}[n]-n_{J}^{\sigma{'}}\right), \label{EOM_q}
\end{equation}
which are exact for a continuous time integration, where the integration time step $\delta t \rightarrow 0$.
The atomic positions and charge degrees of freedom can then be integrated with the same method as for the ground state \cite{ANiklasson09,PSteneteg10,GZheng11,ANiklasson21b}, i.e. the leapfrog velocity Verlet integration of the nuclear coordinates and 
the modified velocity verlet integration scheme for the extended spin-charge degrees of freedom, i.e.\
\begin{equation}
    n_{I}^{\sigma} (t+\delta t) = 2 n_{I}^{\sigma}(t) - n_{I}^{\sigma}(t-\delta t) + \delta t^{2} \Ddot{n}_{I}^{\sigma} (t) + \alpha \sum_{k=0}^{K_{\rm max}} C_{k} n_{I}^{\sigma} (t - k \delta t). \label{n_Int}
\end{equation}
The choice of the $C_{k}$, $\alpha$, and $\kappa = \delta t^{2} \omega^{2}$ are given in Ref.\ \cite{ANiklasson09,PSteneteg10}.

\subsection{Efficient construction of inverse Jacobian kernel} \label{efficient construction}

In a combined matrix-vector notation, the equations of motion in Eq.\ (\ref{EOM_q}) for the electronic degrees of freedom reads,
\begin{equation}
    \ddot{{\textbf{n}}} = -\omega^{2} {\textbf{K}}({\textbf{q}}[{\textbf{n}}]-{\textbf{n}}). \label{n_ddot}
\end{equation}
Here we use a composite notation including both spin indices,  where ${\bf n} = [ \{n^\alpha\}_{i = 1}^N, \{n^\beta\}_{i = 1}^N]$.
The inverse Jacobian matrix, ${\textbf{K}}={\textbf{J}}^{-1}$, acts to keep this propagated spin-resolved charge density very close to the minimum of the shadow potential and the exact $\Delta$SCF excited state. It is critical to avoid calculating and inverting the full Jacobian, which would be very expensive for larger systems.
To avoid this, we can make use of an approximate kernel, ${\textbf{K}}_{0} \approx {\textbf{J}}^{-1}$, as a preconditioner, and we rewrite Eq.\ (\ref{n_ddot}) as
\begin{equation}
    \ddot{{\textbf{n}}} = -\omega^{2}(\textbf{K}_{0}\textbf{J})^{-1} {\textbf{K}}_{0}({\textbf{q}}[{\textbf{n}}]-{\textbf{n}}).
\end{equation}
Rather than determining the Jacobian by computing the derivatives of the residual, ${\bf f}({\bf n}) = {\bf q}[{\bf n}] - {\bf n}$, with respect to the components of $\textbf{n}$, we can instead calculate the directional derivatives, ${\bf f}_{{\bf v}_i}({\bf n})$, for some complete and linearly independent set of vectors, $\{{\bf v}_i\}$. These directional derivatives can be obtained via quantum perturbation theory,  \cite{ANiklasson20,ANiklasson20b} where
\begin{equation}
    \textbf{f}_{\textbf{v}_{i}}(\textbf{n}) \equiv \left.\frac{d \textbf{f}(\textbf{n}+\lambda\textbf{v}_{i})}{d\lambda}\right \vert_{\lambda=0} = \left.\frac{d \textbf{q}[\textbf{n}+\lambda \textbf{v}_{i}]}{d \lambda}\right \vert_{\lambda=0} - \textbf{v}_{i} = {\bf J}{\bf v}_i.
\end{equation}
We can then expand the Jacobian in terms of these directional derivatives, as
\begin{equation}
    \textbf{J} = \sum_{k,l}^N \textbf{f}_{\textbf{v}_{k}} (\textbf{n}){L}_{kl} \textbf{v}^{T}_{l},
\end{equation}
where $\textbf{L}=\textbf{O}^{-1}$, the inverse of the overlap matrix given by ${O}_{kl}=\textbf{v}^{T}_{k}\textbf{v}_{l}$. With the matrix-vector product notation,
\begin{equation}
    \tilde{\textbf{f}}_{\textbf{v}_{i}}(\textbf{n}) = \textbf{K}_{0}\textbf{f}_{\textbf{v}_{i}}(\textbf{n}),
\end{equation}
and
\begin{equation}
    \tilde{\textbf{f}}(\textbf{n}) = \textbf{K}_{0}\textbf{f}(\textbf{n}),
\end{equation}
the preconditioned Jacobian can then be written as
\begin{equation}
    \textbf{K}_{0}\textbf{J} = \sum_{k,l}^N \tilde{\textbf{f}}_{\textbf{v}_{k}}L_{kl} \textbf{v}^{T}_{l}.
\end{equation}
The inverse, i.e.\ the preconditioned kernel, is then given by
\begin{equation}
    (\textbf{K}_{0}\textbf{J})^{-1} = \sum_{k,l}^N \textbf{v}_{k} M_{kl} \tilde{\textbf{f}}^{T}_{\textbf{v}_{l}}.
\end{equation}
Here $\textbf{M}=\tilde{\textbf{O}}^{-1}$ with $\tilde{O}_{ij} = \tilde{\textbf{f}}_{i}^{T} \tilde{\textbf{f}}_{j}$. To profit from this alternative formulation of the inverse preconditioned Jacobian, we use a low-rank approximation of the action of the preconditioned kernel on the residual vector, where 
\begin{equation} \label{low rank kernel}
     (\textbf{K}_{0}\textbf{J})^{-1} \approx \sum_{k,l}^{m} \textbf{v}_{k} M_{kl} \tilde{\textbf{f}}^{T}_{\textbf{v}_{l}}, ~~m < N.
\end{equation}
The vectors $\{\textbf{v}_{k}\}$ are obtained from the Krylov subspace (and their orthonormal Arnoldi complements), \cite{ANiklasson20} i.e.\
\begin{equation}
    \{ \textbf{v}_{k} \} \in \rm{span}^{\perp} \big{\{} \tilde{\textbf{f}}(\textbf{n}), (\textbf{K}_{0}\textbf{J})\tilde{\textbf{f}}(\textbf{n}),(\textbf{K}_{0}\textbf{J})^{2}\tilde{\textbf{f}}(\textbf{n}),... \big{\}}.
\end{equation}
Returning to the electronic equation of motion we obtain,
\begin{equation}
    \ddot{\textbf{n}} \approx - \omega^{2} \sum_{k,l}^{m < N} \textbf{v}_{k} M_{kl} \tilde{\textbf{f}}^{T}_{\textbf{v}_{l}} \textbf{K}_{0}(\textbf{q}[\textbf{n}]-\textbf{n}),
\end{equation}
which is integrated as in Eq.\ (\ref{n_Int}), following Ref.\ \cite{ANiklasson20}. For a given approximation to $\ddot{{\bf n}}_{m}$, we estimate the residual error as 
\begin{equation}
    {\bf r}_{m} = {\bf K}_{\rm 0} ({\bf q}[{\bf n}]-{\bf{n}})-(\sum_{k,l}^{m}\tilde{ \textbf{f}}_{k} M_{kl} \tilde{\textbf{f}}^{T}_{\textbf{v}_{l}}){\bf K}_{\rm 0} ({\bf q}[{\bf n}]-{\bf{n}}).\label{ResidalError}
\end{equation}
and take the relative error 

\begin{equation}
    \tilde{r}_{m} = \frac{||{\bf K}_{\rm 0} ({\bf q}[{\bf n}]-{\bf{n}})-(\sum_{k,l}^{m}\tilde{ \textbf{f}}_{k} M_{kl} \tilde{\textbf{f}}^{T}_{\textbf{v}_{l}}){\bf K}_{\rm 0} ({\bf q}[{\bf n}]-{\bf{n}})||}{||{\bf K}_{\rm 0} ({\bf q}[{\bf n}]-{\bf{n}})||}
\end{equation}

This normalized error estimate essentially measures the accuracy of the resolution of the identity acting on the preconditioned residual function. We take the rank-$m$ approximation of the preconditioned kernel to be converged when this error measure has dropped below a predefined threshold. This is usually achieved in 1 or 2 rank-1 updates. 

\subsection{SCF acceleration} \label{scf acceleration}

As discussed in Refs. \cite{ANiklasson20,VGavini23}, the low-rank approximation for the Jacobian outlined in the previous section can also be used to accelerate SCF convergence in a quasi-Newton approach. In the context of SCC-DFTB, we can write an update rule for the charges using a Newton step,
\begin{equation} \label{qn newton}
{\bf n}_{\rm new} = {\bf n}_{\rm old} - {\bf K}\left({\bf q}[{\bf n}_{\rm old}] - {\bf n}_{\rm old} \right) \approx {\bf n}_{\rm old} - 
\sum_{k,l}^{m < N} \textbf{v}_{k} M_{kl} \tilde{\textbf{f}}^{T}_{\textbf{v}_{l}} \textbf{K}_{0}(\textbf{q}[\textbf{n}_{\rm old}]-\textbf{n}_{\rm old}).
\end{equation}
Here the last term is obtained by inserting the low-rank approximation to the preconditioned kernel given in Eq.\ (\ref{low rank kernel}). The same Newton scheme (or quasi-Newton when a low-rank approximation for the preconditioned Jacobian is used)  can be constructed for $\Delta$SCF convergence, where the occupation number constraints are enforced at the beginning of each iteration, either by simply setting the targeted orbitals to be occupied or unoccupied as desired, or by employing the IMOM scheme. The remainder of the iteration consists of one diagonalization of the charge-dependent Hamiltonian calculated for the new updated charges ${\bf q}[{\bf n}]$, construction of the low-rank kernel approximation, and then updating the SCF charges according to Eq.\ (\ref{qn newton}). This process is repeated until sufficient convergence is reached. The full kernel is computed toward the beginning of the SCF optimization, after several iterations of linear mixing (equivalent to a scaled delta function approximation for the Jacobian) are performed in order to obtain a good starting point, and used as a preconditioner for later iterations. The subsequent iterations are performed using the adaptive preconditioned low-rank approximation to the kernel. Pseudocode for the preconditioned low-rank Jacobian accelerated $\Delta$DFTB scheme is shown in Algorithm \ref{qn delta scf}. One potential advantage of this scheme is that the full kernel preconditioner can be reused across multiple MD timesteps, lowering the computational cost. We use this quasi-Newton acceleration scheme for all of the $\Delta$SCF calculations performed below. The full kernel preconditioner is recomputed in the first SCF cycle of each MD timestep in order to ensure maximum integration stability. The cost of the regular $\Delta$SCF ESMD simulations is reported in the number of Hamiltonian diagonalizations (one for each spin in each SCF iteration) as well as the total number of rank-1 updates in the kernel approximation. For the shadow ESMD simulations, we reuse the full kernel preconditioner across multiple MD timesteps, only recomputing it when the adaptive low-rank protocol fails to reach the desired error tolerance before reaching a max rank of 8. \cite{ANiklasson20} Pseudocode of algorithms for the regular SCC-$\Delta$DFTB and Shadow variants of ESMD are shown in Algorithms \ref{regular esmd} and \ref{shadow esmd}. 

We note that the protocol for the preconditioned quasi-Newton $\Delta$SCF acceleration scheme outlined here is likely not optimal and further improvements should be possible. More sophisticated $\Delta$SCF algorithms would probably result in improved robustness.
Further work exploring how this preconditioned quasi-Newton $\Delta$SCF acceleration scheme can be best designed for regular ESMD is warranted. 

\begin{algorithm}
    \caption{Pseudocode for the quasi-Newton SCC-$\Delta$DFTB Acceleration Scheme. The main difference between this algorithm and a conventional SCF algorithm is in line 11, where the charge update step is performed using a low-rank approximation to the inverse Jacobian kernel. As discussed in Ref. \cite{ANiklasson20}, this scheme is not expected to outperform conventional SCF algorithms (such as DIIS \cite{Pulay82}). However, for $\Delta$SCF where convergence is often difficult to achieve even when the initial guess is good, this quasi-Newton scheme offers reasonable performance.} \label{qn delta scf}
    \begin{algorithmic}[1]
        \State Construct or get initial guess (ES density, $\bf{D}_{\rm \sigma}$, and charge, $\bf{q}$)
        \State Set $\epsilon_{\rm SCF} = 1.0, {\rm ITER} = 1$
        \While{$\epsilon_{\rm SCF} > {\rm SCF_{thresh}}$}
            \State Set $\bf{n}_{\rm old} = q$
            \State Update $\bf H_{\rm \sigma} [n_{\rm old}]$
            \State Obtain $\bf q[n_{\rm old}]$ from diagonalizing $\bf Z^{\rm T}H_{\rm \sigma}Z$ enforcing occupation constraints (Swap HOMO/LUMO or IMOM)
            \If{$\rm ITER == 1$}
                \State Construct full kernel preconditioner $\bf K_{\rm 0} = J^{\rm -1}$
            \EndIf
            \State Construct adaptive rank-m approx $(\bf K_{\rm 0}J)^{\rm -1} \approx \sum_{k,l}^{m < N} \textbf{v}_{k} M_{kl} \tilde{\textbf{f}}^{T}_{\textbf{v}_{l}}$
            \State Update $\bf n_{\rm new} = {\bf n}_{\rm old} - 
(K_{0}J)^{-1} \textbf{K}_{0}(\textbf{q}[\textbf{n}_{\rm old}]-\textbf{n}_{\rm old})$ 
            \State Set $\rm \epsilon_{\rm SCF} = \sqrt{ \frac{1}{N} |\textbf{q}[\textbf{n}_{\rm old}]-\textbf{n}_{\rm old}|^{2}} $
            \State Set $\rm ITER = ITER + 1$
        \EndWhile
    \end{algorithmic}
\end{algorithm}

\begin{algorithm}
    \caption{Pseudocode for the regular SCC-$\Delta$DFTB ESMD simulations. This is essentially a normal direct BOMD algorithm, except that $\Delta$DFTB is used for the electronic degrees of freedom. The nuclear coordinates and velocities are propagated with leapfrog velocity Verlet. Each timestep requires a full SCC-$\Delta$DFTB optimization, which is performed using the method outlined in Algorithm \ref{qn delta scf}. } \label{regular esmd}
    \begin{algorithmic}[1]
        \State Initialize nuclear degrees of freedom: ${\bf m_{\rm I}}$, ${\bf R_{\rm I}}$
        \State Ground state $\bf q_{\rm min}^{\rm GS}$ from SCC-DFTB
        \State Excited state $\bf q_{\rm new} = [{\bf q}_\alpha, {\bf q}_\beta]$ from SCC-$\Delta$DFTB
        \State Obtain exact $U({\bf R})$ and ${\bf F_{\rm I}}$ from $\bf q_{\rm min}$
        \State Initialize velocities, $\bf V_{\rm I}$
        \State Set $t = t_{\rm 0}$
        \While{$t < t_{\rm max}$}
            \State Set initial guess ${\bf q}_{\rm guess}$
            \State Update ${\bf V}_{\rm I} ={\bf V}_{\rm I} + \frac{\delta t}{2 {\bf m_{\rm I}}}{\bf F_{\rm I}} $
            \State Update ${\bf R}_{\rm I} ={\bf R}_{\rm I} + \delta t {\bf V}_{\rm I} $
            \State Get ${\bf q}_{\rm new} \Leftarrow $ SCC-$\Delta$DFTB[${\bf q}_{\rm guess}$] (Full SCF optimization in Alg.\ \ref{qn delta scf})
            \State Compute new forces, ${\bf F}_{I}[{\bf q}_{\rm new}]$
            \State Update ${\bf V}_{\rm I} = {\bf V}_{\rm I} + \frac{\delta t}{2 {\bf m_{\rm I}}}{\bf F_{\rm I}}$
            \State Update $t = t + \delta t$
        \EndWhile
    \end{algorithmic}
\end{algorithm}

\begin{algorithm}
    \caption{Pseudocode for our XL shadow SCC-$\Delta$DFTB ESMD simulations. The nuclear coordinates and velocities are integrated with leapfrog velocity Verlet. The dynamical electronic degrees of freedom are integrated with the modified Verlet scheme described in Ref.\ \cite{ANiklasson20}. The only full SCC-$\Delta$DFTB optimization required is the initial timestep (lines 3 and 4). In each timestep, $\bf \Delta n$ is formed using the adaptive low-rank approximation to $({\bf K}_{\rm 0} {\bf J})^{-1}$. The optimal charges, ${\bf q}_{\rm min}[{\bf n}]$, that exactly minimize the shadow potential, $\mathcal{U}_{\rm \Delta SCF}^{\rm DFTB}$ is obtained in a single step from ${\bf H}_{\rm \sigma}[{\bf n}]$, rather than a full $\Delta$SCF optimization as in conventional ESMD. From these optimal charges, we can obtain the exact forces (${\bf F}_{\rm I}$ in line 18).}   \label{shadow esmd}
    \begin{algorithmic}[1]
        \State Initialize nuclear degrees of freedom: ${\bf m_{\rm I}}$, ${\bf R_{\rm I}}$
        \State Ground state $\bf q_{\rm min}^{\rm GS}$ from SCC-DFTB
        \State Excited state $\bf q_{\rm new} = [{\bf q_{\alpha}}, {\bf q}_{\beta}]^{\rm T}$ from SCC-$\Delta$DFTB
        \State Obtain exact $U({\bf R})$ and ${\bf F_{\rm I}}$ from $\bf q_{\rm min}$
        \State Initialize excited state charges, ${\bf n}_{\rm I} = {\bf q}_{\rm I}$
        \State Initialize velocities, $\bf V_{\rm I}$
        \State Get initial full kernel preconditioner, ${\bf K}_{\rm 0} = \bf{J}^{\rm -1}$
        \State Set initial $\Delta {\bf n} = -({\bf K}_{\rm 0} {\bf J})^{\rm -1}{\bf K}_{\rm 0} ({\bf q}_{\rm min}[{\bf n}] - {\bf n}) = {\bf 0}$ 
        \State Set $t = t_{\rm 0}$
        \While{$t < t_{\rm max}$}
            \State Integrate electronic degrees of freedom: ${\bf n}_{\rm 0} = 2 {\bf n}_{\rm 1} - {\bf n}_{\rm 2} + \delta t^{2} \omega ^{2} \Delta {\bf n} + \alpha \sum_{l=0}^{k} c_{l} {\bf n}_{1-l}$
            \State Update ${\bf n}_{k} = {\bf n}_{k-1},...,{\bf n}_1 = {\bf n}_0, {\bf n} = {\bf n}_{0}$
            \State Update ${\bf V}_{\rm I} ={\bf V}_{\rm I} + \frac{\delta t}{2 {\bf m_{\rm I}}}{\bf F_{\rm I}} $
            \State Update ${\bf R}_{\rm I} ={\bf R}_{\rm I} + \delta t {\bf V}_{\rm I} $
            \State Construct ${\bf H}_{\sigma}[{\bf n}]$
            \State Obtain ${\bf q}_{\rm min}[{\bf n}] \Leftarrow$ from diagonalization of $\bf Z^{\rm T} H_{\rm \sigma}Z$
            \State Construct $\Delta {\bf n} = -({\bf K}_{\rm 0} {\bf J})^{\rm -1}{\bf K}_{\rm 0} ({\bf q}_{\rm min}[{\bf n}] - {\bf n}) $ with adaptive low-rank approximation
            \State From ${\bf q}_{\rm min}[{\bf n}]$ obtain $\mathcal{U}_{\rm \Delta SCF}^{\rm DFTB}({\bf R},{\bf n})$, and new ${\bf F}_{\rm I}$
            \State Update ${\bf V}_{\rm I} = {\bf V}_{\rm I} + \frac{\delta t}{2 {\bf m_{\rm I}}}{\bf F_{\rm I}}$
            \State Update $t = t + \delta t$
        \EndWhile
    \end{algorithmic}
\end{algorithm}

\section{Results and Discussion}

To evaluate and demonstrate the shadow excited state molecular dynamics with the $\Delta$SCF method we have used the SCC-$\Delta$DFTB implementation described above in all of our examples. The main goal of this paper is to understand how the generalization of the most recent formulation of ground-state XL-BOMD to shadow ESMD can improve the performance and applicability of the $\Delta$SCF method. The regular $\Delta$SCF ESMD simulations, used for comparison, were carried out using the quasi-Newton SCF acceleration scheme described in subsection \ref{scf acceleration}.  In these calculations the full kernel preconditioner was reconstructed in the first $\Delta$SCF cycle of each timestep, and only recomputed when the adaptive low-rank scheme failed to reach an residual error tolerance of ${\bf r}_m = 0.1$ in Eq.\ (\ref{ResidalError}) before reaching rank 8 in the kernel approximation \cite{ANiklasson20}. For the shadow $\Delta$SCF ESMD simulations, the preconditioner was reused across timesteps and only recomputed when the low-rank update protocol failed to reach an error tolerance of 0.5 before rank 8. The only full $\Delta$SCF optimization performed in the shadow ESMD simulations is the initial calculation at time zero. 

 \subsection{Reusing the old density as initial guess}
 
\begin{figure}
        \centering
        \includegraphics[width=0.98\linewidth]{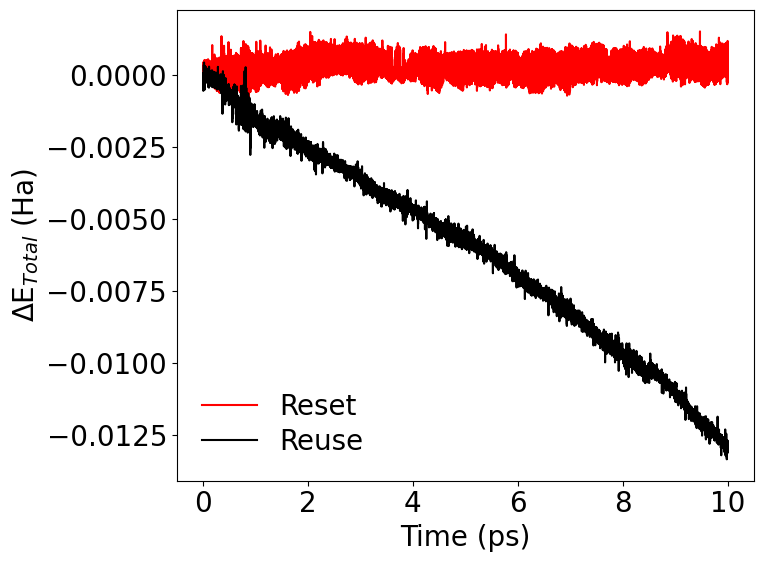}
        \caption{Fluctuations in total energy (kinetic + potential) for the H$_2$CO + H$_2$O $\Delta$SCF ESMD simulation with an integration timestep of 0.1 femtoseconds. An SCF convergence threshold of $\epsilon_{\rm SCF}$ = 10$^{-4}$ Ha was used. Reusing the previous excited state charge solution leads to a significant systematic drift even for this short integration time step. The $\Delta$SCF algorithm uses the quasi-Newton schemed discussed earlier together with non-{\it aufbau} occupation constraints. The lowest triplet excited state corresponding to HOMO to LUMO was targeted. Reset guess refers to $\Delta$SCF with an initial guess constructed from a non-{\it aufbau} occupation of the eigenvectors of the charge-independent Hamiltonian ($H_{0}$), corresponding to a superposition of overlapping charge densities from the neutral atoms,  whereas reuse guess refers to using the converged charge-density from the previous timestep as an initial guess for the current timestep. The accompanying convergence statistics are shown in Tab.\ \ref{tab:reuse}.}
        \label{fig:drift}
    \end{figure}

    As mentioned previously, the simplest approach to provide a good initial guess in $\Delta$SCF ESMD simulations is to use the converged density from the previous timestep. As long as the state of interest continues to be adiabatically separated from it neighbors, this simple guess should work well. Even if the states would swap character from one timestep to the next, it is still possible to construct an improved initial guess if a history of more than one previous state is being propagated. There is a long history of constructing initial guesses using information from previous timesteps, mostly in the context of ground state BOMD \cite{MCPayne92,PPulay04,JMHerbert05,ANiklasson06,TDKuhne07,JFang16,EPolack21,ZAskarpour25}. As a test, we performed SCC-$\Delta$DFTB ESMD on the lowest triplet excited state of a H$_{2}$CO + H$_{2}$O system, either resetting the initial guess each iteration (as a superposition of electron densities of neutral atoms) or reusing the previous iterations converged charge density as an initial guess. Higher-order extrapolation schemes could also be used, but these typically exhibit stability problems unless very tight convergence is enforced and are likely less useful in practice. Our simple test should suffice to demonstrate the utility of reusing old density information, as well as the drawbacks of such a technique. In Table \ref{tab:reuse} we show the convergence statistics for the two approaches. In both cases, the $\Delta$SCF acceleration scheme described in Algorithm \ref{qn delta scf} was used. When the initial guess is 'reset' each timestep, the initial guess is constructed by enforcing a non-\textit{aufbau} occupation of the non-interacting atomic spin densities. By reusing the old density as an initial guess, we can already achieve a decrease in the average number SCF iterations of more than 80$\%$. At the same time, it is well known that naively reusing the old density will lead to a systematic energy drift \cite{DRemler90,PPulay04,ANiklasson06} because of the broken time-reversal symmetry and this effect is seen also in our simulations, as is demonstrated in Fig. \ref{fig:drift}. By tightening or loosening the SCF convergence, the energy drift can be reduced or increased. 
    
    Nonetheless, even this experiment suggests that there is potential for significant performance improvement in terms of SCF iterations for $\Delta$SCF ESMD by incorporating information from the trajectory history, if the problem of systematic energy drift can be avoided. This is the same situation that was encountered in the context of ground state BOMD, and the solution to these two objectives (lower computational expense and elimination of systematic energy drift) has eventually matured into the shadow XL-BOMD method \cite{ANiklasson21b}, in which the electronic density is formally propagated as an additional dynamical variable and the exact BO potential is replaced by an approximate shadow potential that can be exactly solved at low cost in each timestep. We now proceed to describe the performance of the same method applied to ESMD in the context of SCC-$\Delta$DFTB. Because $\Delta$SCF approaches typically suffer from much worse convergence behavior than their ground-state SCF counterparts, we conjectured that the improved stability and elimination of self-consistent field optimizations offered by a shadow ESMD approach within the XL-BOMD formalism would lead to even larger performance gains over regular ESMD simulations than those observed for the ground state.

    \begin{table}[]
        \centering
        \begin{tabular}{|c|c|c|}
        \hline
           Method  & Average SCF Iterations & Maximum SCF Iterations\\
           \hline 
           Reset Guess  &  32.96 & 70 \\
           Reuse Guess & 2.15 & 7 \\
           \hline
        \end{tabular}
        \caption{SCF convergence statistics for a 10 picosecond $\Delta$SCF ESMD simulation on H$_2$CO + H$_2$O with a timestep of 0.1 femtoseconds. For the accompanying total energy fluctuations, see Fig.\ \ref{fig:drift}. An SCF convergence threshold of $\epsilon_{\rm SCF}$ = 10$^{-4}$ Ha was used. The $\Delta$SCF algorithm uses the quasi-Newton scheme discussed earlier (Alg \ref{qn delta scf}) together with non-{\it aufbau} occupation constraints. The lowest triplet excited state corresponding to a HOMO to LUMO excitation was targeted. Reset guess refers to $\Delta$SCF with an initial guess constructed from a non-{\textit{aufbau}} occupation of the eigenvectors of the charge-independent Hamiltonian ($H_{\sigma}^{(0)}$), corresponding to a superposition of overlapping charge densities of neutral atoms, whereas reuse guess refers to using the converged excited charge-density from the previous timestep as an initial guess for the current timestep.}
        \label{tab:reuse}
    \end{table}

    \subsection{Comparison to regular ESMD}

\begin{figure}
            \includegraphics[width=0.98\textwidth]{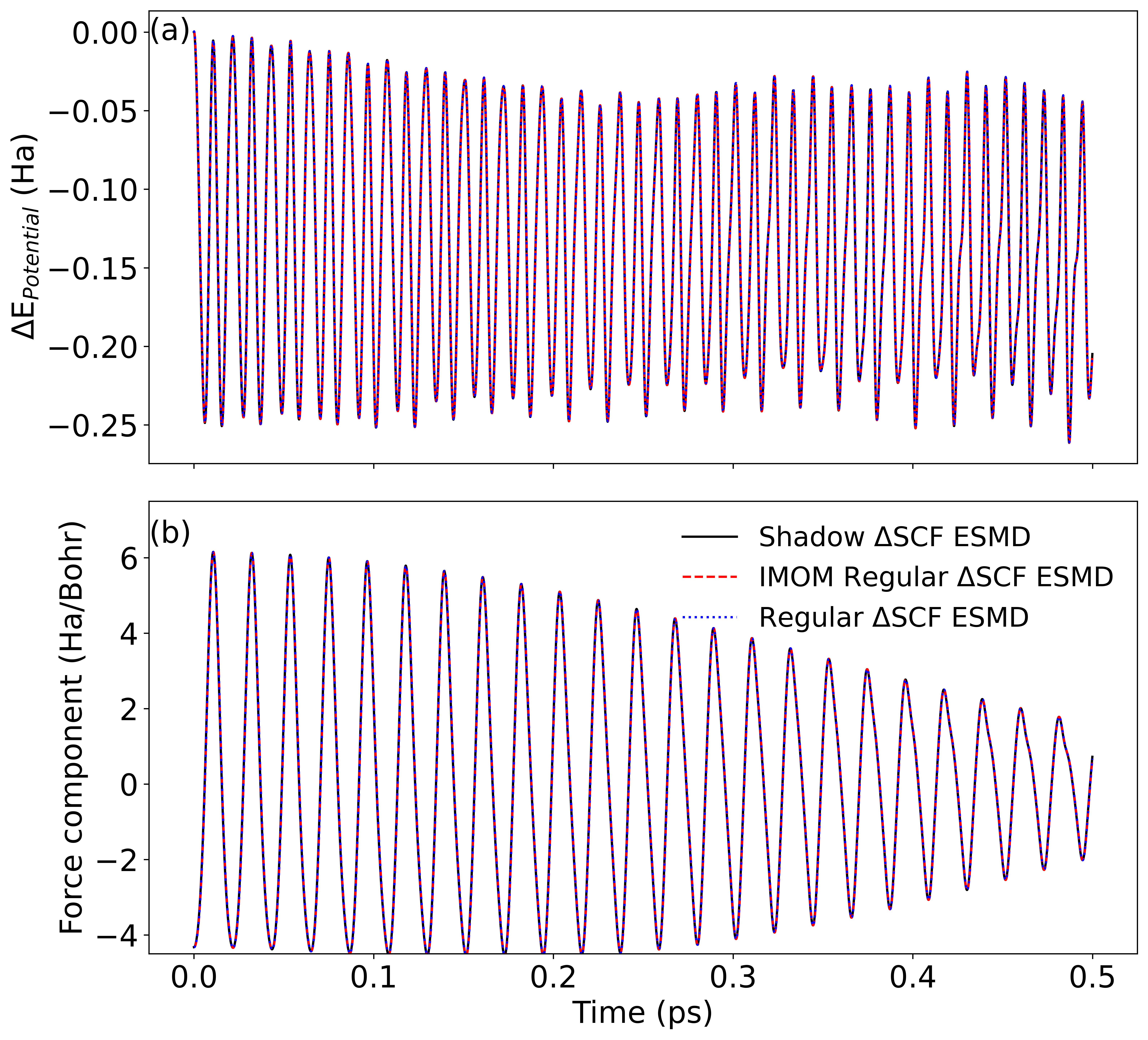}
            \caption{Comparison of exact and shadow potential and forces calculated along the same trajectory (H$_2$O + H$_2$CO). The nuclear trajectory is evolved according the shadow ESMD forces with a timestep of 0.1 femtoseconds. For the conventional $\Delta$SCF calculations, an SCF convergence threshold of 10$^{-4}$ was used. A non-\textit{aufbau} occupation of the non-interacting atomic spin densities was used as the initial guess. The lowest singlet HOMO to LUMO excitation was targeted. In the shadow ESMD, the full preconditioner is constructed at the first timestep and reused throughout the dynamics unless the adaptive low-rank kernel approximation fails to achieve a residual error of 0.5 before rank 8. In the regular $\Delta$SCF ESMD, the full preconditioner is recomputed at the start of each SCF optimization (once per timestep) and reconstructed if the low-rank kernel approximation fails to achieve a residual error of 0.1 before rank 8. The exact and shadow potential energy and forces lie virtually on top of one another. }
        \label{fig:comp}    
    \end{figure}

    We begin by examining the accuracy of the shadow $\Delta$SCF ESMD method relative to regular `exact' $\Delta$SCF ESMD. For a meaningful simulation, it is important that the shadow potential and forces closely follow their exact counterparts. We tested this by simulating the dynamics of the lowest singlet excited state of the same small system as above, i.e.\ one water and one formaldehyde molecule separated by a short distance, with the shadow $\Delta$SCF ESMD method. Then, as a reference, we calculated the exact (fully $\Delta$SCF optimized) Born-Oppenheimer potential energy and forces along the same shadow trajectory. The initial guess was constructed from a non-\textit{aufbau} occupation of the non-interacting atomic spin densities.
    A comparison of the potential energy and x-component of the force on one of the water hydrogen atoms along the simulated molecular trajectory is shown in Fig. \ref{fig:comp}. We compute the 'exact' $\Delta$SCF solution with and without the IMOM technique for enforcing the excited state configuration. Both 'exact' $\Delta$SCF simulations used the quasi-Newton acceleration scheme described earlier in Alg.\ \ref{qn delta scf}. The potential energy curves along the trajectory lie virtually on top of one another for all three methods, as do the force component curves. This indicates that the shadow $\Delta$SCF ESMD produces a very accurate simulation of the underlying exact $\Delta$SCF excited state surface. This is also what is expected from the theory and ground state XL-BOMD simulations \cite{ANiklasson20}. Thus, even if the shadow potential is approximated with a linearization around the propagated electronic degrees of freedom, the potential energy is virtually exact.
    
    Table \ref{tab:perf-sametraj} reports performance statistics for the three methods on the same trajectory. Because the computational expense of $\Delta$SCF ESMD is dominated by the self-consistent solution of the excited state electronic density, improvements in SCF convergence translate directly to improvements in computational cost, where each SCF iteration requires two Hamiltonian diagonalizations (one for each spin). From Table \ref{tab:perf-sametraj} it is immediately evident that the IMOM approach offers a dramatic improvement over the 'brute force' $\Delta$SCF without IMOM, cutting the average number of diagonalizations per time step (or SCF iterations) by an order of magnitude, although the maximum number of diagonalizations (SCF iterations) is only halved from standard $\Delta$SCF (last column of Tab. \ref{tab:perf-sametraj}). This provides some indication that difficult SCF convergence situations can still plague even more sophisticated convergence schemes for $\Delta$SCF. Moving from IMOM $\Delta$SCF to shadow $\Delta$SCF offers a further improvement in both average and worst-case performance. It is important to note that no iterative SCF cycles are performed in the shadow $\Delta$SCF scheme (except for an initial full $\Delta$SCF calculation at time zero), and the number of diagonalizations per time step is simply two (one for each spin), as shown at the bottom row of Tab. \ref{tab:perf-sametraj}. The table also shows the average number of rank-1 updates in the kernel approximation in each time step as well as the number of preconditioners constructed. The average rank-1 updates refers to the adaptive low-rank approximation to the preconditioned kernel, Eq.\  (\ref{low rank kernel}), which is used to construct ${\bf \ddot n}$ in the shadow ESMD and to update ${\bf q}_{\rm new}$ in the quasi-Newton $\Delta$SCF scheme. In general, we need one or two rank-1 updates per timestep in shadow ESMD, and one or two updates per SCF iteration in the quasi-Newton $\Delta$SCF. Several iterations of linear mixing (equivalent to a delta-function for the Jacobian) are performed at the start of the SCF cycle in order to obtain a good starting point for the quasi-Newton $\Delta$SCF scheme. These linear mixing iterations therefore do not involve any low-rank update steps, but still require two diagonalizations per iteration. Because the IMOM ESMD converges very quickly (5-10 iterations per timestep), there are only relatively few quasi-Newton steps, and therefore on average there are more diagonalizations than low-rank updates per timestep. The last column reports the average number of times the full kernel (inverse Jacobian) is constructed per MD timestep. In the shadow approach, the full kernel preconditioner is computed at the beginning of the simulation and then only reconstructed when the adaptive low-rank procedure fails to converge sufficiently quickly. By contrast, in the regular $\Delta$SCF ESMD we rebuild the kernel each timestep. In all, the shadow ESMD approach is at least five times less expensive than even the most affordable regular ESMD simulation (IMOM quasi-Newton $\Delta$SCF ESMD ($\epsilon = 10^{-4}$)), and over an order of magnitude compared to the more tightly converged case. As we will see later, the simulation stability achieved by shadow ESMD is at the same time comparable or greater than regular $\Delta$SCF ESMD with a tight SCF convergence threshold ($\epsilon = 10^{-6}$). This is a result of the improved robustness and stability over time offered by a shadow ESMD approach. This advantage is especially pronounced when the target excited state becomes nearly degenerate with a higher-lying state, as demonstrated in our next example.

\begin{figure}

            \includegraphics[width=0.98\textwidth]{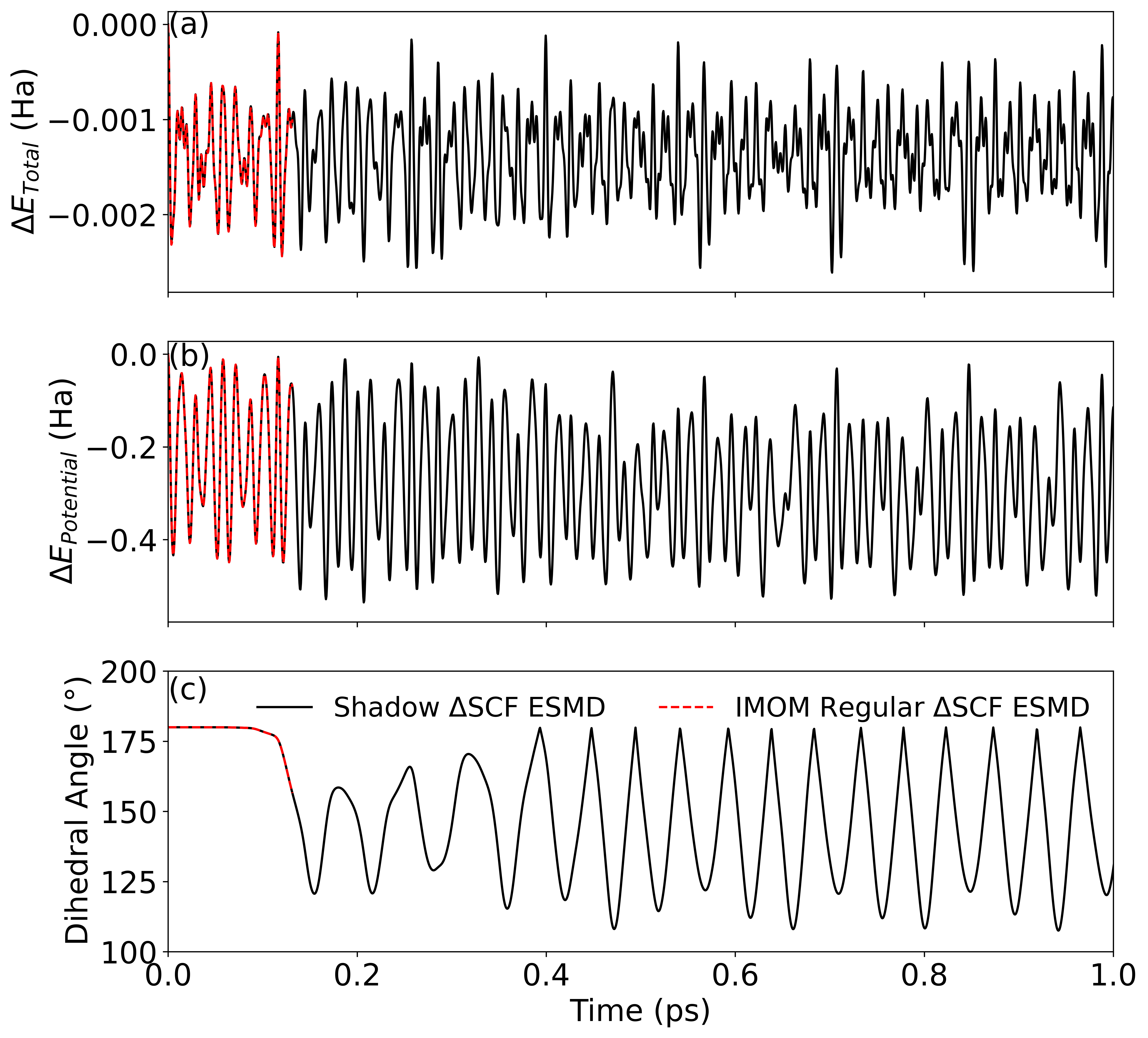}
        \caption{Comparison of exact and shadow ESMD for ethylene with a timestep of 0.5 femtoseconds. Total energy (kinetic + potential) fluctuations (panel a) lie virtually on top of one another until around 150 femtoseconds, at which point the exact IMOM regular ESMD fails to converge, whereas the shadow ESMD continues to follow the original state smoothly. A similar pattern is shown in the fluctuations of the potential energy (b). The convergence difficulties encountered by the IMOM $\Delta$SCF ESMD occur as the dihderal angle begins to deviate from planarity (c). The exact IMOM regular ESMD used an SCF convergence threshold of 10$^{-6}$ Ha. Further performance statistics are shown in Tab.\ \ref{tab:perf-ethylene}.
        }
        \label{fig:ethylene}
    \end{figure}

    In Fig. \ref{fig:ethylene} we compare the total energy, potential energy, and H$_1$-C$_2$-C$_4$-H$_6$ dihedral angle of ethylene in its lowest singlet excited state calculated with shadow $\Delta$SCF ESMD and with 'exact' IMOM regular $\Delta$SCF ESMD, using a fairly tight convergence threshold (10$^{-6}$). As for the previous simulation, the initial guess is constructed from a non-\textit{aufbau} occupation of the non-interacting atomic spin density. The fluctuations in the total and potential energy along the trajectory (panels (a) and (b) panel of Fig. \ref{fig:ethylene}) lie virtually on top of one another up until about 150 femtoseconds, further indicating that the shadow ESMD simulation is accompanied by essentially no loss in accuracy. The dihedral angles (panel (c) of Fig. \ref{fig:ethylene}) for the regular $\Delta$SCF and shadow $\Delta$SCF ESMD methods are also virtually on top of each other. However, at around 150 femtoseconds, the 'exact' IMOM regular $\Delta$SCF ESMD trajectory hits a step where the excited state density can no longer be converged, occurring as the dihedral angle begins to deviate significantly from planarity. By contrast, the shadow $\Delta$SCF ESMD is able to continue the smooth propagation of the original state without any stability problems. Because the shadow potential and forces very closely mirror the underlying exact surface, and because the inverse Jacobian kernel keeps the dynamical electronic charge degrees of freedom propagating very tightly about the optimal shadow charges, the shadow ESMD results in a dynamics that very closely follows the targeted state. This example demonstrates the power of the shadow $\Delta$SCF ESMD approach in the context of extended Lagrangian dynamics for systems where SCF convergence can be particularly difficult. By using the approximate shadow potential and propagating the excited state density itself as a dynamical variable, we are able to improve not just stability, but also robustness in the sense of state tracking. This feature may be of particular importance in excited state dynamics simulations of extended systems where the density of states is high and many trivially unavoided crossings may exist. \cite{fernandez-alberti_identification_2012,meek_evaluation_2014,wang_simple_2014}

    For the same reason, we anticipate that the shadow $\Delta$SCF ESMD approach should also avoid problems of variational collapse to the ground state in the course of the dynamics, which can be a significant concern in direct $\Delta$SCF ESMD simulations. To see this differently, one can view the shadow ESMD- much like shadow XL-BOMD for the ground state- as a constrained SCF optimization over time. By including in the shadow potential a harmonic oscillator centered about the optimal charges, ${\bf q}_{\rm min}[{\bf n}]$, the targeted excited state is now at least a local minimum of the shadow potential surface, whereas the same state may correspond only to a saddle-point on the underlying 'exact' surface. \cite{burton_energy_2022} Hence, as long as the initial 'exact' (fully $\Delta$SCF optimized) electronic solution corresponds to the desired excited state and as long as the adiabatic limit where $\omega$ is large holds, then the constrained optimization over time will converge to the targeted excited state. In this sense, shadow ESMD is a powerful technique for finding state-specific solutions because it changes the underlying potential surface. This bears at least some similarity to the squared gradient minimization method for targeting excited states, \cite{hait_excited_2020,hait_orbital_2021} which also works by changing the optimization surface, converting saddle points into minima of the target function. Another way to understand the stability and robustness of the shadow ESMD simulation is to consider the hypothetical situation that the excited state density over time converges to the ground state. At that instant the optimized excited state determining the shadow potential would be orthogonal to the ground state. A collapse to the ground state is therefore prevented. The single-step optimization of the shadow potential also naturally avoids oscillations between different states that can occur in a regular SCF optimization scheme.

    \subsection{Performance comparison}

    In Tab.\ \ref{tab:perf-ethylene} we report the performance statistics for the shadow $\Delta$SCF ESMD and regular $\Delta$SCF ESMD (with and without IMOM) for ethylene. We again observe a marked improvement in average performance from the regular quasi-Newton scheme to IMOM, and again marked improvements from IMOM to shadow $\Delta$SCF ESMD. Since both of the regular $\Delta$SCF algorithms reach a point where convergence fails (see Fig. \ref{fig:ethylene}), we only list the maximum number of diagonalizations (SCF iterations) for the timesteps which do converge. In the case of ethylene, we find that shadow ESMD requires 4 or 10 times fewer diagonalizations than regular IMOM $\Delta$SCF ESMD, depending on the convergence threshold. Without using IMOM, the regular $\Delta$SCF ESMD is 13 or 26 times more expensive in terms of diagonalizations, again depending on the convergence threshold. The shadow $\Delta$SCF ESMD thus offers an accurate propagation of the exact excited state surface while improving the average computational cost by a factor of 4-10 even over sophisticated $\Delta$SCF convergence methods, and enables smooth propagation through regions of near degeneracy between electronic states that prove catastrophic for standard $\Delta$SCF ESMD. Using a looser $\Delta$SCF convergence threshold ($\epsilon = 10^{-4}$) lowers the computational cost of conventional ESMD simulations substantially (still more than 4 times slower than shadow ESMD), but can lead to much worse stability than the shadow ESMD approach, as we will see later. It is possible that even more sophisticated (and likely more expensive) $\Delta$SCF algorithms could also continue the smooth propagation of the state of interest in ethylene, but it is still noteworthy that shadow $\Delta$SCF ESMD ameliorates these issues while at the same time dramatically lowering computational cost.

\begin{figure}
        \centering
        \includegraphics[width=0.98\linewidth]{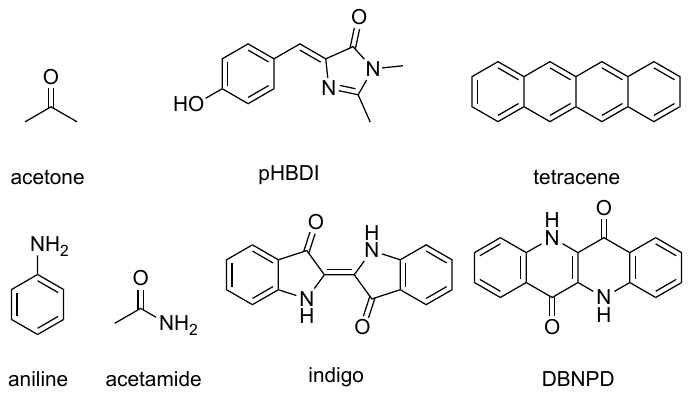}
        \caption{Several small organic molecules used to study the performance of the Shadow ESMD method compared to IMOM $\Delta$SCF dynamics. The initial geometries were taken from \cite{kowalczyk_self-consistent_2016}. Performance results for these systems are found in Tab. \ref{tab:more-perf}}
        \label{fig:molecules}
    \end{figure}

    We went on to examine the performance gain offered by shadow $\Delta$SCF ESMD over regular IMOM $\Delta$SCF ESMD for several systems, consisting of small and medium sized organic molecules. The initial geometries were taken from Ref \cite{kowalczyk_self-consistent_2016} and the structures are shown in Fig. \ref{fig:molecules}. The performance statistics are displayed in Tab. \ref{tab:more-perf}. We used two SCF convergence thresholds for the IMOM ESMD calculations: 10$^{-4}$ Ha and 10$^{-6}$ Ha. When the looser threshold is used, the average SCF performance is in general improved by a factor of 2-3 compared to regular $\Delta$SCF ESMD with a tight convergence threshold. However, as shown in Fig. \ref{fig:comp-dt1}, regular $\Delta$SCF ESMD with a looser SCF threshold displays much lower stability than shadow $\Delta$SCF ESMD. Even at the looser convergence threshold, regular ESMD generally requires about 5 times more diagonalizations than shadow ESMD on average. As seen in Fig. \ref{fig:comp-dt1}, using a tighter SCF convergence threshold (10$^{-6}$) leads to comparable stability to shadow $\Delta$SCF ESMD- in the absence of convergence failures which can occur in the regular $\Delta$SCF ESMD and completely prevent further simulation. When such a tight convergence threshold is used, the shadow ESMD is usually about 10 to 15 times cheaper in terms of diagonalizations. This indicates the 'best of both worlds' offered by shadow $\Delta$SCF ESMD over regular $\Delta$SCF ESMD, i.e.\ better stability than regular $\Delta$SCF ESMD with a loose convergence threshold, and better computational performance than regular $\Delta$ SCF ESMD with a tight convergence threshold.

    \begin{table}[]
        \centering
        \small
        \begin{tabular}{|c|c|c|c|c|}
        \hline
        Method     & Avg. Diag./timestep & Avg. Rank-1 & Avg. Full ${\bf K_{0}}$ &Max. Diag.\\
            &  & Updates/timestep &  Builds/timestep  &\\

        \hline
        QN $\Delta$SCF ESMD ($\epsilon = 10^{-4}$)    & 79.58 & 20.47 & 1.00   &96\\
        QN $\Delta$SCF ESMD ($\epsilon = 10^{-6}$) & 102.89 & 46.71 & 1.00 &132\\
        IMOM/QN $\Delta$SCF ESMD ($\epsilon = 10^{-4}$) & 7.11 & 2.87 & 1.00  &12\\
        IMOM/QN $\Delta$SCF ESMD ($\epsilon = 10^{-6}$) & 22.06 & 11.51 & 1.00  &40\\    
        Shadow $\Delta$SCF ESMD  & 2 & 1.02 & 0.0002 &2\\
        \hline
        \end{tabular}
        \caption{ Performance statistics for ESMD trajectory of H$_2$O + H$_2$CO with a timestep of 0.1 femtoseconds with regular $\Delta$DFTB (with and without IMOM) and shadow ESMD. The nuclear trajectory is evolved according the shadow ESMD forces with an integration timestep of 0.1 femtoseconds. 
        }
        \label{tab:perf-sametraj}
    \end{table}

    \begin{table}[]
        \centering
        \small
        \begin{tabular}{|c|c|c|c|c|}
        \hline
        Method     & Avg. Diag./timestep & Avg. Rank-1 \ & Avg. Full ${\bf K_{0}}$  &Max. Diag.$^a$\\
              &  & Updates/timestep &  Builds/timestep  &\\
        \hline
        QN $\Delta$SCF ESMD ($\epsilon = 10^{-4}$)    & 21.88 & 10.86 & 1.00   &198\\
        QN $\Delta$SCF ESMD ($\epsilon = 10^{-6}$) & 42.20 & 27.31 & 1.00  &528\\
        IMOM/QN $\Delta$SCF ESMD ($\epsilon = 10^{-4}$) & 4.64 & 1.40 & 1.00  &12\\
        IMOM/QN $\Delta$SCF ESMD ($\epsilon = 10^{-6}$) & 15.96 & 7.18 & 1.00  &134\\    
        Shadow $\Delta$SCF ESMD  & 2.00 & 1.00 & 0.0003  &2\\
        \hline
        \end{tabular}
        \caption{Performance statistics for ESMD trajectories of ethylene with a timestep of 0.5 femtoseconds with brute force regular $\Delta$SCF (with and without IMOM) and shadow ESMD. In practice, each low-rank update has a cost of about one half of an SCF iteration. The exact ESMD used an error tolerance of 0.1 for the low-rank Jacobian update. The shadow (XL-BOMD) used an error tolerance of 0.5.}
        $^a$ Maximum diagonalizations for timesteps where the SCF converges.
        \label{tab:perf-ethylene}
    \end{table}

    \begin{table}[]
        \centering
        \begin{adjustbox}{width=1.2\textwidth,center=\textwidth}
        \small
        \begin{tabular}{|c|c|c|c|c|c|c|}
        \hline
            & $\Delta$SCF ($\epsilon$ = 10$^{-4}$)  & Avg. Rank-1 & $\Delta$SCF ($\epsilon$ = 10$^{-6}$) &Avg. Rank-1 & Shadow ESMD &  Avg. Rank-1\\
            System & Avg. Diag. /timestep & updates/timestep & Avg. Diag. /timestep &  updates/timestep & Diag. /timestep& updates/timestep \\
            \hline
           acetone  & 13.06 & 5.02 & 31.19 & 18.32 & 2.00 & 1.05 \\
            acetamide & 22.17 & 13.11 & 48.25 & 34.17 & 2.00 & 1.07\\
            aniline & 11.92 & 2.98 & 36.62 & 15.59 & 2.00 & 1.01  \\
            pHBDI & 11.02 & 2.65 & 31.11 & 14.56 & 2.00 & 1.00\\
            tetracene & 7.99 & 1.00 & 23.00 & 8.83 & 2.00 & 1.00\\
            DBNPD & 10.00 & 2.04 & 23.93 & 9.76 & 2.00 & 1.00 \\
            indigo & 10.16 & 2.13 & 30.08 & 12.82 & 2.00 & 1.00 \\
            \hline
        \end{tabular}
        \end{adjustbox}
        \caption{Performance statistics for the set of small molecules shown in \ref{fig:molecules}, using IMOM $\Delta$SCF with two SCF thresholds, and shadow ESMD. The timestep was 0.5 femtoseconds for each simulation, and 2 picoseconds of dynamics were run for each system.
        The regular $\Delta$SCF ESMD simulations of acetamide eventually reached a point where SCF convergence could not be reached (with both thresholds), similar to the situation for ethylene.}
        \label{tab:more-perf}
    \end{table}

    \subsection{Stability of shadow ESMD}

\begin{figure}
            \centering
            \includegraphics[width=0.98\linewidth]{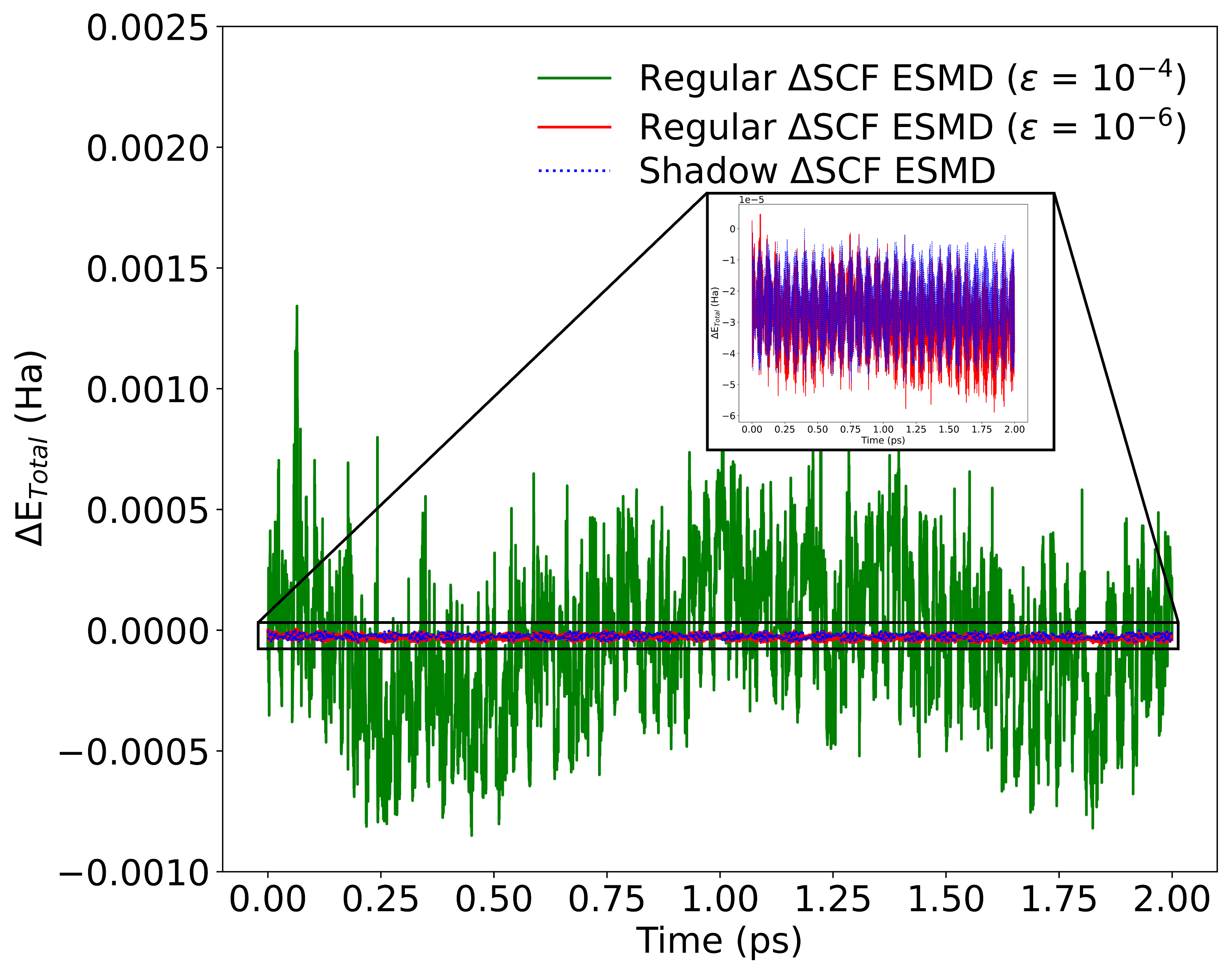}
            \caption{Fluctuations in total energy with timestep dt = 0.1 femtoseconds for ESMD simulations of the first singlet excited state of aniline, calculated with exact (regular BOMD) $\Delta$SCF and Shadow $\Delta$SCF. The regular $\Delta$SCF simulations used SCF convergence thresholds of 10$^{-4}$ and 10$^{-6}$ Ha. The inset shows a close up of the total energy fluctuations for Shadow $\Delta$SCF and regular $\Delta$SCF with a convergence threshold of 10$^{-6}$ Ha.
            The fluctuations in the total energy of the more tightly converged regular $\Delta$SCF ESMD and the shadow $\Delta$SCF ESMD are virtually on top of each other. At this convergence threshold, regular $\Delta$SCF ESMD is 18 times more expensive than shadow ESMD, in terms of diagonalizations alone. Further performance statistics are shown in Tab.\ \ref{tab:more-perf}.
            }
            \label{fig:comp-dt1}
        \end{figure}

 In Fig. \ref{fig:comp-dt1} we demonstrate the enhanced stability of shadow ESMD relative to exact (regular $\Delta$SCF) dynamics, when the same timestep is used. Comparable stability can only be obtained by employing very tight SCF convergence thresholds, but the computational expense increases significantly (see Tab.\ \ref{tab:more-perf}). Using a looser convergence threshold can lessen the performance gap between 'exact' $\Delta$SCF dynamics and shadow ESMD, but leads to a substantial loss of stability, as seen in Fig.\ \ref{fig:comp-dt1}. This suggests that shadow ESMD may allow somewhat larger time steps than exact regular $\Delta$SCF dynamics techniques with loose convergence thresholds, potentially leading to further computational savings. 

\begin{figure}
            \includegraphics[width=0.98\textwidth]{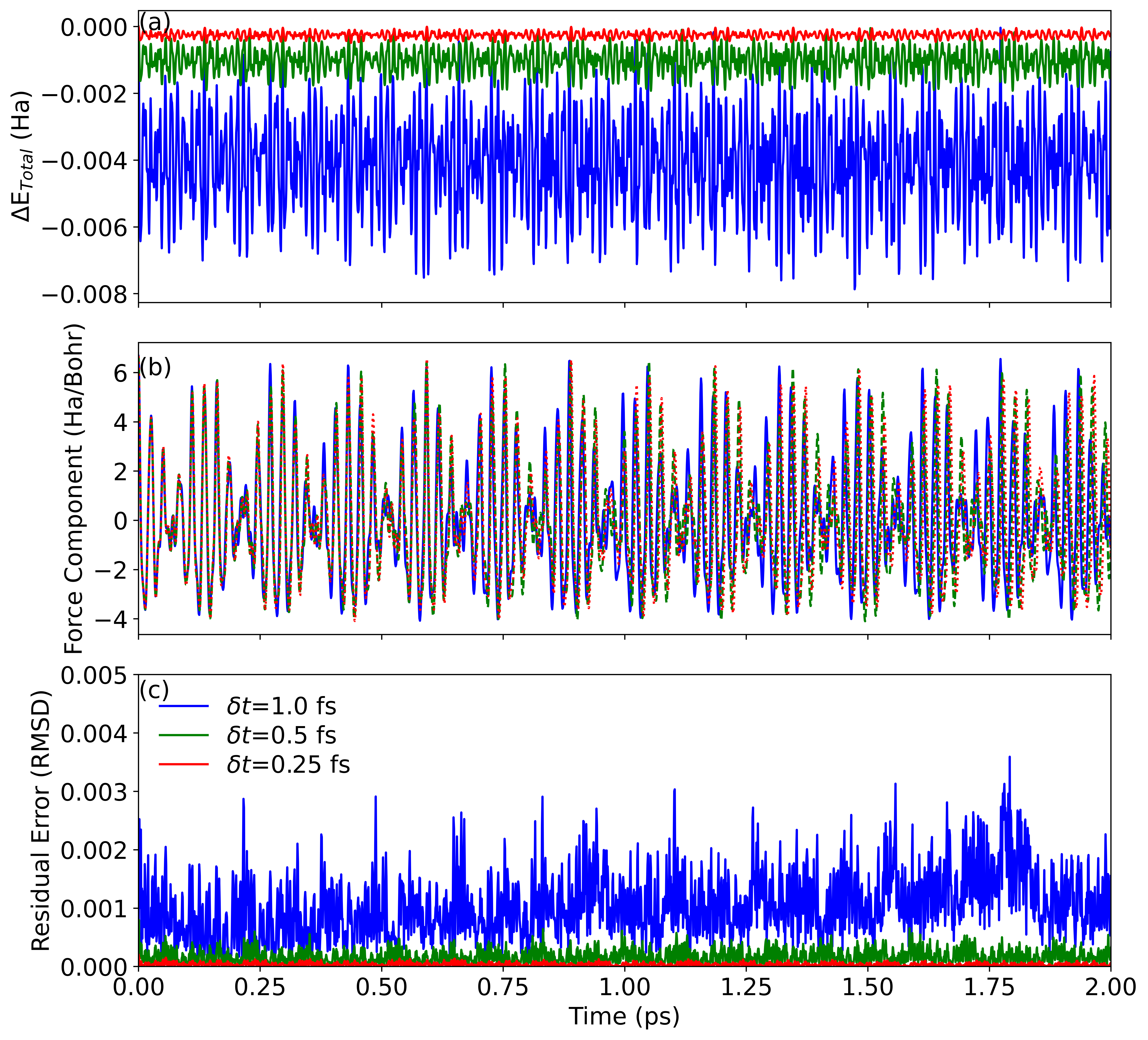}
  
        \caption{Fluctuations in total energy (kinetic + potential), x-component of the force on one of the ethylene hydrogens, and residual error (RMS deviation of total charge density) for various timesteps in shadow $\Delta$SCF simulations of ethylene. 
        By halving the timestep, we reduce the fluctuations in energy and residual error by a factor of four, while avoiding any noticeable change in the forces.}
        \label{fig:stability}
    \end{figure}

    As for ground state XL-BOMD simulations, halving the timestep should lead to a fourfold decrease in the fluctuations of the total energy and residual error (here defined as the RMS error of the total charge density) also for the excited state. This $\delta t^{2}$ scaling is demonstrated in Fig. \ref{fig:stability} for timesteps of 1, 0.5, and 0.25 femtoseconds for shadow ESMD simulations of ethylene. The characteristic improvement of energy and residual fluctuations with decreasing $\delta t$ allows us to tune the stability of the shadow ESMD dynamics. A timestep of 1 femtosecond leads to energy fluctuations which are borderline acceptable (a few milliHartree) for a shadow ESMD simulation and no systematic energy drift is observed. At the same time, increasing the timestep shows very little influence on the forces, as shown in panel (b) of Fig. \ref{fig:stability}. Hence, tuning the size of the integration time step in the simulation does not affect the accuracy of the propagated forces. The improved stability offered by shadow ESMD may thus allow larger integration time steps to be used than those required by regular 'exact' ESMD (unless a very tight convergence threshold is used), particularly where there is a high density of electronic excited states or when the state of interest may undergo an unwanted change of character.

\section{Conclusions}

    We have introduced a shadow $\Delta$SCF ESMD approach for KS-DFT based on the XL-BOMD technique. The method was implemented and evaluated using SCC-$\Delta$DFTB theory. Our results suggest that a shadow $\Delta$SCF ESMD can transfer the reduction in computational cost and the improved stability observed for shadow ground state XL-BOMD to the excited state. In fact, the performance gain for the shadow $\Delta$SCF ESMD simulations seems to be even greater than for the ground state, because the excited state $\Delta$SCF optimization is typically far more challenging. Hence, the shadow potential approximation and the propagation of the electronic density as an additional extended dynamical variable provides a far cheaper, potentially more robust alternative to reconverging the exact excited state density at each timestep. In some cases where the conventional $\Delta$DFTB is not able to converge to the desired excited state, shadow $\Delta$SCF ESMD is able to continue to smoothly propagate through phase space. 
    
    It will be interesting to combine the shadow ESMD approach with nonadiabatic dynamics simulations (e.g.\ Tully's surface hopping) and to explore its potential for multi-surface phenomena. In that context, it may be necessary to develop a multi-state shadow $\Delta$SCF, which we expect would be a widely useful method for non-adiabatic MD (NAMD) simulations. The combination of a multi-surface $\Delta$SCF method with a shadow ESMD could resolve long-standing difficulties with the use of complete active space methods in NAMD, namely unwanted orbital rotations that change the active space composition. Much work has gone into state-tracking methods for excited state dynamics \cite{marie_excited_2023,zaitsevskii_discontinuities_1994,saade_excited_2024,tran_tracking_2019,Zhanserkeev21} but the shadow ESMD may offer a better approach for directly propagating the desired electronic densities.

    We also note that the SCC-$\Delta$DFTB method can be reparameterized to offer better accuracy for electronic excited states and that our current implementation does not attempt to achieve high accuracy for the systems under study. A more accurate propagation could likely be achieved by a higher-quality parameterization and this would be an interesting further direction to pursue, especially in combination with modern machine-learning methods. Additionally, the shadow ESMD method should be tested with higher levels of theory, such as $\Delta$SCF with Hartree-Fock and first-principles KS-DFT. The results we have presented here, based on the SCC-DFTB theory, suggest that these may be fruitful directions for the affordable and stable simulation of ESMD at a variety of levels of theory. It is also possible to combine the shadow ESMD scheme with higher-level shadow potentials and integration techniques for the dynamical electronic degrees of freedom, following the same approach as for ground state shadow XL-BOMD. \cite{Niklasson23} This would lead to even greater stability improvements at a marginal increase in computational cost.

\section*{Acknowledgments}
This work is supported by the U.S. Department of Energy, Office of Basic Energy Sciences (FWP LANLE8AN) and by the U.S. Department of Energy through the Los Alamos National Laboratory. Los Alamos National Laboratory is operated by Triad National Security, LLC, for the National Nuclear Security Administration of the U.S. Department of Energy Contract No. 892333218NCA000001.
OJF is supported by the U.S. Department of Energy, Office of Science, Office of Advanced Scientific Computing Research, Department of Energy Computational Science Graduate Fellowship under Award Number DE-SC0023112 and the AMOS program of the U.S. Department of Energy, Office of Science, Basic Energy Sciences, Chemical Sciences, Geosciences, and Biosciences Division.

\section*{Data Availability}
The data that supports the findings of this study are available within the article and its supplementary material. 

\section*{Supporting Information}
Additional data demonstrating the effect of various choices for initial guess in regular ESMD simulations on average SCF convergence. All data and plots for making the figures can be found at 10.5281/zenodo.16423745.

\newpage
\bibliography{references/mndo_new_xy-2}

\begin{thebibliography}{100}

\bibitem{levine_isomerization_2007}
B.~G. Levine and T.~J. Martínez,
\newblock Annual Review of Physical Chemistry {\bf 58}, 613 (2007).

\bibitem{AkimovRev}
A.~V. Akimov, A.~J. Neukirch, and O.~V. Prezhdo,
\newblock Chemical Reviews {\bf 113}, 4496 (2013),
\newblock PMID: 23627277.

\bibitem{hammes-schiffer_theory_2010}
S.~Hammes-Schiffer and A.~A. Stuchebrukhov,
\newblock Chemical Reviews {\bf 110}, 6939 (2010).

\bibitem{wodtke__electronically_2004}
A.~M. Wodtke, J.~C. Tully, and D.~J. Auerbach,
\newblock International Reviews in Physical Chemistry {\bf 23}, 513 (2004).

\bibitem{mukherjee_beyond_2019}
B.~Mukherjee, K.~Naskar, S.~Mukherjee, S.~Ghosh, T.~Sahoo, and S.~Adhikari,
\newblock International Reviews in Physical Chemistry {\bf 38}, 287 (2019).

\bibitem{frank_molecular_1998}
I.~Frank, J.~Hutter, D.~Marx, and M.~Parrinello,
\newblock The Journal of Chemical Physics {\bf 108}, 4060 (1998).

\bibitem{birgisson_decoherence_2025}
B.~O. Birgisson, A.~O. Dohn, H.~Jónsson, and G.~Levi,
\newblock The Journal of Chemical Physics {\bf 162}, 044306 (2025).

\bibitem{JCTully1971}
J.~C. Tully and R.~K. Preston,
\newblock The Journal of Chemical Physics {\bf 55}, 562 (1971).

\bibitem{JCTully12}
J.~Tully,
\newblock J. Chem. Phys. {\bf 137}, 22A301 (2012).

\bibitem{li_ab_2005}
X.~Li, J.~C. Tully, H.~B. Schlegel, and M.~J. Frisch,
\newblock The Journal of Chemical Physics {\bf 123}, 084106 (2005).

\bibitem{ben-nun_ab_2000}
M.~Ben-Nun, J.~Quenneville, and T.~J. Martínez,
\newblock The Journal of Physical Chemistry A {\bf 104}, 5161 (2000).

\bibitem{prigogine_ab_2002}
M.~Ben‐Nun and T.~J. Martínez,
\newblock \textit{{Ab} {Initio}} {Quantum} {Molecular} {Dynamics},
\newblock in {\em Advances in {Chemical} {Physics}}, edited by I.~Prigogine and S.~A. Rice, volume 121, pages 439--512, Wiley, 1 edition, 2002.

\bibitem{curchod_ab_2018}
B.~F.~E. Curchod and T.~J. Martínez,
\newblock Chemical Reviews {\bf 118}, 3305 (2018).

\bibitem{crespo-otero_recent_2018}
R.~Crespo-Otero and M.~Barbatti,
\newblock Chemical Reviews {\bf 118}, 7026 (2018).

\bibitem{makhov_ab_2014}
D.~V. Makhov, W.~J. Glover, T.~J. Martinez, and D.~V. Shalashilin,
\newblock The Journal of Chemical Physics {\bf 141}, 054110 (2014).

\bibitem{baeck_ab_2003}
K.~K. Baeck and T.~J. Martinez,
\newblock Chemical Physics Letters {\bf 375}, 299 (2003).

\bibitem{choi_ab_2004}
H.~Choi, K.~K. Baeck, and T.~J. Martinez,
\newblock Chemical Physics Letters {\bf 398}, 407 (2004).

\bibitem{hait_prediction_2024}
D.~Hait, D.~Lahana, O.~J. Fajen, A.~S.~P. Paz, P.~A. Unzueta, B.~Rana, L.~Lu, Y.~Wang, E.~F. Kjønstad, H.~Koch, and T.~J. Martínez,
\newblock The Journal of Chemical Physics {\bf 160}, 244101 (2024).

\bibitem{kjonstad_photoinduced_2024}
E.~F. Kjønstad, O.~J. Fajen, A.~C. Paul, S.~Angelico, D.~Mayer, M.~Gühr, T.~J.~A. Wolf, T.~J. Martínez, and H.~Koch,
\newblock Nature Communications {\bf 15}, 10128 (2024).

\bibitem{coe_ab_2007}
J.~D. Coe, B.~G. Levine, and T.~J. Martínez,
\newblock The Journal of Physical Chemistry A {\bf 111}, 11302 (2007).

\bibitem{park_caspt2_2017}
J.~W. Park and T.~Shiozaki,
\newblock Journal of Chemical Theory and Computation {\bf 13}, 3676 (2017).

\bibitem{polyak_ultrafast_2019}
I.~Polyak, L.~Hutton, R.~Crespo-Otero, M.~Barbatti, and P.~J. Knowles,
\newblock Journal of Chemical Theory and Computation {\bf 15}, 3929 (2019).

\bibitem{hohenstein_tensor_2012}
E.~G. Hohenstein, R.~M. Parrish, and T.~J. Martínez,
\newblock The Journal of Chemical Physics {\bf 137}, 044103 (2012).

\bibitem{hohenstein_quartic_2013}
E.~G. Hohenstein, S.~I.~L. Kokkila, R.~M. Parrish, and T.~J. Martínez,
\newblock The Journal of Chemical Physics {\bf 138}, 124111 (2013).

\bibitem{parrish_tensor_2012}
R.~M. Parrish, E.~G. Hohenstein, T.~J. Martínez, and C.~D. Sherrill,
\newblock The Journal of Chemical Physics {\bf 137}, 224106 (2012).

\bibitem{hohenstein_tensor_2013}
E.~G. Hohenstein, S.~I.~L. Kokkila, R.~M. Parrish, and T.~J. Martínez,
\newblock The Journal of Physical Chemistry B {\bf 117}, 12972 (2013).

\bibitem{hohenstein_rank_2019}
E.~G. Hohenstein, Y.~Zhao, R.~M. Parrish, and T.~J. Martínez,
\newblock The Journal of Chemical Physics {\bf 151}, 164121 (2019).

\bibitem{hohenstein_rank-reduced_2022}
E.~G. Hohenstein, B.~S. Fales, R.~M. Parrish, and T.~J. Martínez,
\newblock The Journal of Chemical Physics {\bf 156}, 054102 (2022).

\bibitem{song_reduced_2018}
C.~Song and T.~J. Martínez,
\newblock The Journal of Chemical Physics {\bf 149}, 044108 (2018).

\bibitem{fales_large-scale_2018}
B.~S. Fales, S.~Seritan, N.~F. Settje, B.~G. Levine, H.~Koch, and T.~J. Martínez,
\newblock Journal of Chemical Theory and Computation {\bf 14}, 4139 (2018).

\bibitem{casida_chapter}
M.~E. Casida,
\newblock {\em Time-Dependent Density Functional Response Theory for Molecules}, chapter~5, pages 155--192,
\newblock 2005.

\bibitem{casida_molecular_1998}
M.~E. Casida, C.~Jamorski, K.~C. Casida, and D.~R. Salahub,
\newblock The Journal of Chemical Physics {\bf 108}, 4439 (1998).

\bibitem{hirata_time-dependent_1999}
S.~Hirata and M.~Head-Gordon,
\newblock Chemical Physics Letters {\bf 314}, 291 (1999).

\bibitem{TNelson20}
T.~R. Nelson, A.~J. White, J.~A. Bjorgaard, A.~E. Sifain, Y.~Zhang, B.~Nebgen, S.~Fernandez-Alberti, D.~Mozyrsky, A.~E. Roitberg, and S.~Tretiak,
\newblock Chemical Reviews {\bf 120}, 2215 (2020),
\newblock PMID: 32040312.

\bibitem{roos1980}
B.~O. Roos,
\newblock International Journal of Quantum Chemistry {\bf 18}, 175 (1980).

\bibitem{krylov_size_consistent_2001}
A.~I. Krylov,
\newblock Chemical Physics Letters {\bf 338}, 375 (2001).

\bibitem{lee_eliminating_2018}
S.~Lee, M.~Filatov, S.~Lee, and C.~H. Choi,
\newblock The Journal of Chemical Physics {\bf 149}, 104101 (2018).

\bibitem{lee_efficient_2019}
S.~Lee, E.~E. Kim, H.~Nakata, S.~Lee, and C.~H. Choi,
\newblock The Journal of Chemical Physics {\bf 150}, 184111 (2019).

\bibitem{van_aggelen_exchange-correlation_2013}
H.~Van~Aggelen, Y.~Yang, and W.~Yang,
\newblock Physical Review A {\bf 88}, 030501 (2013).

\bibitem{yang_double_2013}
Y.~Yang, H.~Van~Aggelen, and W.~Yang,
\newblock The Journal of Chemical Physics {\bf 139}, 224105 (2013).

\bibitem{bannwarth_holehole_2020}
C.~Bannwarth, J.~K. Yu, E.~G. Hohenstein, and T.~J. Martínez,
\newblock The Journal of Chemical Physics {\bf 153}, 024110 (2020).

\bibitem{Bagus65}
P.~S. Bagus,
\newblock Phys. Rev. {\bf 139}, A619 (1965).

\bibitem{Slater1972}
J.~C. Slater,
\newblock Statistical exchange-correlation in the self-consistent field,
\newblock volume~6 of {\em Advances in Quantum Chemistry}, pages 1--92, Academic Press, 1972.

\bibitem{OGunnarsson76}
O.~Gunnarsson and B.~I. Lundqvist,
\newblock Phys. Rev. B {\bf 13}, 4274 (1976).

\bibitem{Gunnarsson89}
R.~O. Jones and O.~Gunnarsson,
\newblock Rev. Mod. Phys. {\bf 61}, 689 (1989).

\bibitem{BILundqvist04}
A.~Hellman, B.~Razaznejad, and B.~I. Lundqvist,
\newblock The Journal of Chemical Physics {\bf 120}, 4593 (2004).

\bibitem{Gavnholt08}
J.~Gavnholt, T.~Olsen, M.~Engelund, and J.~Schi\o{}tz,
\newblock Phys. Rev. B {\bf 78}, 075441 (2008).

\bibitem{TZiegler09}
T.~Ziegler, M.~Seth, M.~Krykunov, J.~Autschbach, and F.~Wang,
\newblock The Journal of Chemical Physics {\bf 130}, 154102 (2009).

\bibitem{TKowalczyk11}
T.~Kowalczyk, S.~R. Yost, and T.~V. Voorhis,
\newblock The Journal of Chemical Physics {\bf 134}, 054128 (2011).

\bibitem{RMaurer11}
R.~J. Maurer and K.~Reuter,
\newblock The Journal of Chemical Physics {\bf 135}, 224303 (2011).

\bibitem{TKowalczyk16}
T.~Kowalczyk, K.~Le, and S.~Irle,
\newblock Journal of Chemical Theory and Computation {\bf 12}, 313 (2016),
\newblock PMID: 26587877.

\bibitem{WMackrodt18}
W.~C. Mackrodt, S.~Salustro, B.~Civalleri, and R.~Dovesi,
\newblock Journal of Physics: Condensed Matter {\bf 30}, 495901 (2018).

\bibitem{Hait2020}
D.~Hait and M.~Head-Gordon,
\newblock Journal of Chemical Theory and Computation {\bf 16}, 1699 (2020),
\newblock PMID: 32017554.

\bibitem{DHait21}
D.~Hait and M.~Head-Gordon,
\newblock The Journal of Physical Chemistry Letters {\bf 12}, 4517 (2021),
\newblock PMID: 33961437.

\bibitem{EVandaele22}
E.~Vandaele, M.~Mališ, and S.~Luber,
\newblock The Journal of Chemical Physics {\bf 156}, 130901 (2022).

\bibitem{TKowalczyk23}
M.~Y. Deshaye, A.~T. Wrede, and T.~Kowalczyk,
\newblock The Journal of Chemical Physics {\bf 158}, 134104 (2023).

\bibitem{JKahk23}
J.~M. Kahk and J.~Lischner,
\newblock Journal of Chemical Theory and Computation {\bf 19}, 3276 (2023),
\newblock PMID: 37163299.

\bibitem{WYang24}
W.~Yang and P.~W. Ayers,
\newblock Foundation for the delta-scf approach in density functional theory, 2024.

\bibitem{HPham25}
H.~D.~M. Pham and R.~Z. Khaliullin,
\newblock Journal of Chemical Theory and Computation {\bf 21}, 3902 (2025),
\newblock PMID: 40175288.

\bibitem{gilbert_self-consistent_2008}
A.~T.~B. Gilbert, N.~A. Besley, and P.~M.~W. Gill,
\newblock The Journal of Physical Chemistry A {\bf 112}, 13164 (2008).

\bibitem{hait_orbital_2021}
D.~Hait and M.~Head-Gordon,
\newblock The Journal of Physical Chemistry Letters {\bf 12}, 4517 (2021).

\bibitem{PrezhdoPRL}
C.~F. Craig, W.~R. Duncan, and O.~V. Prezhdo,
\newblock Phys. Rev. Lett. {\bf 95}, 163001 (2005).

\bibitem{pyxaid}
A.~V. Akimov and O.~V. Prezhdo,
\newblock Journal of Chemical Theory and Computation {\bf 9}, 4959 (2013),
\newblock PMID: 26583414.

\bibitem{neukirch}
A.~J. Neukirch, L.~C. Shamberger, E.~Abad, B.~J. Haycock, H.~Wang, J.~Ortega, O.~V. Prezhdo, and J.~P. Lewis,
\newblock Journal of Chemical Theory and Computation {\bf 10}, 14 (2014),
\newblock PMID: 26579888.

\bibitem{RaquelEP}
R.~Esteban-Puyuelo and B.~Sanyal,
\newblock Phys. Rev. B {\bf 103}, 235433 (2021).

\bibitem{MCPayne92}
M.~C. Payne, M.~P. Teter, D.~C. Allan, T.~A. Arias, and J.~D. Joannopoulos,
\newblock Rev. Mod. Phys. {\bf 64}, 1045 (1992).

\bibitem{PPulay04}
P.~Pulay and G.~Fogarasi,
\newblock Chem. Phys. Lett. {\bf 386}, 272 (2004).

\bibitem{JMHerbert05}
J.~Herbert and M.~Head-Gordon,
\newblock Phys. Chem. Chem. Phys. {\bf 7}, 3269 (2005).

\bibitem{ANiklasson06}
A.~M.~N. Niklasson, C.~J. Tymczak, and M.~Challacombe,
\newblock Phys. Rev. Lett. {\bf 97}, 123001 (2006).

\bibitem{TDKuhne07}
T.~D. K\"{u}hne, M.~Krack, F.~R. Mohamed, and M.~Parrinello,
\newblock Phys. Rev. Lett. {\bf 98}, 066401 (2007).

\bibitem{JFang16}
J.~Fang, X.~Gao, H.~Song, and H.~Wang,
\newblock J. Chem. Phys {\bf 144}, 244103 (2016).

\bibitem{EPolack21}
{\'E}.~Polack, G.~Dusson, B.~Stamm, and F.~Lipparini,
\newblock Journal of Chemical Theory and Computation {\bf 17}, 6965 (2021),
\newblock PMID: 34623810.

\bibitem{DRemler90}
D.~K. Remler and P.~A. Madden,
\newblock Mol.\ Phys. {\bf 70}, 921 (1990).

\bibitem{RCar85}
R.~Car and M.~Parrinello,
\newblock Phys. Rev. Lett. {\bf 55}, 2471 (1985).

\bibitem{JHutter12}
J.~Hutter,
\newblock WIREs Comput. Mol. Sci. {\bf 2}, 604 (2012).

\bibitem{ANiklasson08}
A.~M.~N. Niklasson,
\newblock Phys. Rev. Lett. {\bf 100}, 123004 (2008).

\bibitem{ANiklasson17}
A.~M.~N. Niklasson,
\newblock J. Chem. Phys. {\bf 147}, 054103 (2017).

\bibitem{ANiklasson20}
A.~M.~N. Niklasson,
\newblock J. Chem. Phys. {\bf 152}, 104103 (2020).

\bibitem{ANiklasson21b}
A.~M.~N. Niklasson,
\newblock Eur. Phys. J. B {\bf 94}, 164 (2021).

\bibitem{Niklasson23}
A.~M.~N. Niklasson and C.~F.~A. Negre,
\newblock The Journal of Chemical Physics {\bf 158}, 154105 (2023).

\bibitem{HYoshida90}
H.~Yoshida,
\newblock Phys. Lett. A {\bf 150}, 262 (1990).

\bibitem{CGrebogi90}
C.~Grebogi, S.~M. Hammel, J.~A. Yorke, and T.~Saur,
\newblock Phys. Rev. Lett. {\bf 65}, 1527 (1990).

\bibitem{SToxvaerd94}
S.~Toxvaerd,
\newblock Phys. Rev. E {\bf 50}, 2271 (1994).

\bibitem{JGans00}
J.~Gans and D.~Shalloway,
\newblock Phys. Rev. E {\bf 61}, 4587 (2000).

\bibitem{BJLeimkuhler94}
B.~J. Leimkuhler and R.~D. Skeel,
\newblock J. Comput. Phys. {\bf 112}, 117 (1994).

\bibitem{RDEngel05}
R.~D. Engel, R.~D. Skeel, and M.~Drees,
\newblock J. Comput. Phys. {\bf 206}, 432 (2005).

\bibitem{BJLeimkuhler04}
B.~Leimkuhler and S.~Reich,
\newblock {\em Simulating Hamiltonian Dynamics},
\newblock Cambridge University Press, Cambridge, 2004.

\bibitem{ShadowHamiltonian}
S.~D. Bond and B.~J. Leimkuhler,
\newblock {\em Molecular dynamics and the accuracy of numerically computed averages},
\newblock Cambride University Press, United Kingdom, 2007.

\bibitem{SToxvaerd12}
S.~Toxvaerd, O.~J. Heilmann, and J.~C. Dyre,
\newblock J. Chem. Phys. {\bf 136}, 224106 (2012).

\bibitem{KDHammonds21}
K.~D. Hammonds and D.~M. Heyes,
\newblock J. Chem. Phys. {\bf 154}, 174102 (2021).

\bibitem{CNegre22}
C.~F.~A. Negre, M.~E. Wall, and A.~M.~N. Niklasson,
\newblock Graph-based quantum response theory and shadow born-oppenheimer molecular dynamics, 2022.

\bibitem{JBjorgaard18}
J.~A. Bjorgaard, D.~Sheppard, S.~Tretiak, and A.~M.~N. Niklasson,
\newblock Journal of Chemical Theory and Computation {\bf 14}, 799 (2018),
\newblock PMID: 29316401.

\bibitem{MElstner98}
M.~Elstner, D.~Poresag, G.~Jungnickel, J.~Elsner, M.~Haugk, T.~Frauenheim, S.~Suhai, and G.~Seifert,
\newblock Phys. Rev. B {\bf 58}, 7260 (1998).

\bibitem{MFinnis98}
M.~W. Finnis, A.~T. Paxton, M.~Methfessel, and M.~van Schilfgarde,
\newblock Phys. Rev. Lett. {\bf 81}, 5149 (1998).

\bibitem{TFrauenheim00}
T.~Frauenheim, G.~Seifert, M.~Elstner, Z.~Hajnal, G.~Jungnickel, D.~Poresag, S.~Suhai, and R.~Scholz,
\newblock Phys. Stat. sol. {\bf 217}, 41 (2000).

\bibitem{hohen}
P.~Hohenberg and W.~Kohn,
\newblock Phys. Rev. {\bf 136}, B:864 (1964).

\bibitem{KohnSham65}
W.~Kohn and L.~J. Sham,
\newblock Phys. Rev. {\bf 140}, 1133 (1965).

\bibitem{Barth_1972}
U.~von Barth and L.~Hedin,
\newblock Journal of Physics C: Solid State Physics {\bf 5}, 1629 (1972).

\bibitem{RMDreizler90}
R.~Dreizler and K.~Gross,
\newblock {\em Density-functional theory},
\newblock Springer Verlag, Berlin Heidelberg, 1990.

\bibitem{RWentzcovitch92}
R.~M. Wentzcovitch, J.~L. Martins, and P.~B. Allen,
\newblock Phys. Rev. B {\bf 45}, R11372 (1992).

\bibitem{RParr89}
R.~G. Parr and W.~Yang,
\newblock {\em Density-functional theory of atoms and molecules},
\newblock Oxford University Press, Oxford, 1989.

\bibitem{EMartinez15}
E.~Martinez, M.~J. Cawkwell, A.~F. Voter, and A.~M.~N. Niklasson,
\newblock J. Chem. Phys. {\bf 142}, 154120 (2015).

\bibitem{ziegler_implementation_2012}
T.~Ziegler, M.~Krykunov, and J.~Cullen,
\newblock The Journal of Chemical Physics {\bf 136}, 124107 (2012).

\bibitem{bourne_worster_reliable_2021}
S.~Bourne~Worster, O.~Feighan, and F.~R. Manby,
\newblock The Journal of Chemical Physics {\bf 154}, 124106 (2021).

\bibitem{zheng_implementation_2009}
G.~Zheng, M.~Lundberg, J.~Jakowski, T.~Vreven, M.~J. Frisch, and K.~Morokuma,
\newblock International Journal of Quantum Chemistry {\bf 109}, 1841 (2009).

\bibitem{carter-fenk_state-targeted_2020}
K.~Carter-Fenk and J.~M. Herbert,
\newblock Journal of Chemical Theory and Computation {\bf 16}, 5067 (2020).

\bibitem{ivanov_method_2021}
A.~V. Ivanov, G.~Levi, E.~{\"O}. Jónsson, and H.~Jónsson,
\newblock Journal of Chemical Theory and Computation {\bf 17}, 5034 (2021).

\bibitem{schmerwitz_variational_2022}
Y.~L.~A. Schmerwitz, A.~V. Ivanov, E.~{\"O"}. Jónsson, H.~Jónsson, and G.~Levi,
\newblock The Journal of Physical Chemistry Letters {\bf 13}, 3990 (2022).

\bibitem{schmerwitz_calculations_2023}
Y.~L.~A. Schmerwitz, G.~Levi, and H.~Jónsson,
\newblock Journal of Chemical Theory and Computation {\bf 19}, 3634 (2023).

\bibitem{schmerwitz_saddle_2024}
Y.~L.~A. Schmerwitz, N.~Urgell~Ollé, G.~Levi, and H.~Jónsson,
\newblock Saddle {Point} {Search} {Algorithms} for {Variational} {Density} {Functional} {Calculations} of {Excited} {Electronic} {States} with {Self}-{Interaction} {Correction},
\newblock in {\em Proceedings of the {Platform} for {Advanced} {Scientific} {Computing} {Conference}}, pages 1--11, Zurich Switzerland, 2024, ACM.

\bibitem{sigurdarson_orbital-optimized_2023}
A.~E. Sigurdarson, Y.~L.~A. Schmerwitz, D.~K.~V. Tveiten, G.~Levi, and H.~Jónsson,
\newblock The Journal of Chemical Physics {\bf 159}, 214109 (2023).

\bibitem{hait_excited_2020}
D.~Hait and M.~Head-Gordon,
\newblock Journal of Chemical Theory and Computation {\bf 16}, 1699 (2020).

\bibitem{GLevi20}
G.~Levi, A.~V. Ivanov, and H.~Jónsson,
\newblock Journal of Chemical Theory and Computation {\bf 16}, 6968 (2020),
\newblock PMID: 33064484.

\bibitem{ANiklasson09}
A.~M.~N. Niklasson, P.~Steneteg, A.~Odell, N.~Bock, M.~Challacombe, C.~J. Tymczak, E.~Holmstrom, G.~Zheng, and V.~Weber,
\newblock J. Chem. Phys. {\bf 130}, 214109 (2009).

\bibitem{PSteneteg10}
P.~Steneteg, I.~A. Abrikosov, V.~Weber, and A.~M.~N. Niklasson,
\newblock Phys. Rev. B {\bf 82}, 075110 (2010).

\bibitem{GZheng11}
G.~Zheng, A.~M.~N. Niklasson, and M.~Karplus,
\newblock J. Chem. Phys. {\bf 135}, 044122 (2011).

\bibitem{MFoulkes89}
W.~M.~C. Foulkes and R.~Haydock,
\newblock Phys. Rev. B {\bf 39}, 12520 (1989).

\bibitem{DPorezag95}
D.~Porezag, T.~Frauenheim, T.~K\"ohler, G.~Seifert, and R.~Kaschner,
\newblock Phys. Rev. B {\bf 51}, 12947 (1995).

\bibitem{BAradi07}
B.~Aradi, B.~Hourahine, and T.~Frauenheim,
\newblock The Journal of Physical Chemistry A {\bf 111}, 5678 (2007).

\bibitem{BHourahine20}
B.~H. et~al.,
\newblock J. Chem. Phys. {\bf 152}, 124101 (2020).

\bibitem{PDral2015}
P.~O. Dral, O.~A. von Lilienfeld, and W.~Thiel,
\newblock Journal of Chemical Theory and Computation {\bf 11}, 2120 (2015),
\newblock PMID: 26146493.

\bibitem{DYaron18}
H.~Li, C.~Collins, M.~Tanha, G.~J. Gordon, and D.~J. Yaron,
\newblock Journal of Chemical Theory and Computation {\bf 14}, 5764 (2018),
\newblock PMID: 30351008.

\bibitem{PDral20}
P.~O. Dral,
\newblock The Journal of Physical Chemistry Letters {\bf 11}, 2336 (2020),
\newblock PMID: 32125858.

\bibitem{ZGuoqing22}
G.~Zhou, N.~Lubbers, K.~Barros, S.~Tretiak, and B.~Nebgen,
\newblock Proc. Nat. Ac. Sci. {\bf 119}, 2120333119 (2022).

\bibitem{DYaron23}
F.~Hu, F.~He, and D.~J. Yaron,
\newblock Semiempirical hamiltonians learned from data can have accuracy comparable to density functional theory, 2023.

\bibitem{LATTE}
N.~Bock, M.~J. Cawkwell, J.~D. Coe, A.~Krishnapriyan, M.~P. Kroonblawd, A.~Lang, , C.~Liu, E.~M. Saez, S.~M. Mniszewski, C.~F.~A. Negre, A.~M.~N. Niklasson, E.~Sanville, M.~A. Wood, and P.~Yang,
\newblock {\sc {LATTE}}, 2008,
\newblock \mbox{L}os Alamos National Laboratory (LA- CC-10004), http://www.github.com/lanl/latte.

\bibitem{MCawkwell12}
M.~J. Cawkwell and A.~M.~N. Niklasson,
\newblock J. Chem. Phys. {\bf 137}, 134105 (2012).

\bibitem{AKrishnapriyan17}
A.~Krishnapryian, P.~Yang, A.~M.~N. Niklasson, and M.~J. Cawkwell,
\newblock J. Chem. Theory Comput. {\bf 13}, 6191 (2017).

\bibitem{melix_spin_2016}
P.~Melix, A.~F. Oliveira, R.~Rüger, and T.~Heine,
\newblock Theoretical Chemistry Accounts {\bf 135}, 232 (2016).

\bibitem{VQVuong23}
V.-Q. Vuong, B.~Aradi, A.~M.~N. Niklasson, Q.~Cui, and S.~Irle,
\newblock Journal of Chemical Theory and Computation {\bf 19}, 7592 (2023),
\newblock PMID: 37890454.

\bibitem{RKLindsey17}
R.~K. Lindsey, L.~E. Fried, and N.~Goldman,
\newblock Journal of Chemical Theory and Computation {\bf 13}, 6222 (2017),
\newblock PMID: 29113430.

\bibitem{NGoldman23}
N.~Goldman, L.~E. Fried, R.~K. Lindsey, C.~H. Pham, and R.~Dettori,
\newblock The Journal of Chemical Physics {\bf 158}, 144112 (2023).

\bibitem{ANiklasson20b}
A.~M.~N. Niklasson,
\newblock J. Chem. Theory Comput. {\bf 16}, 3628 (2020).

\bibitem{VGavini23}
S.~Das and V.~Gavini,
\newblock Phys. Rev. B {\bf 107}, 125133 (2023).

\bibitem{Pulay82}
P.~Pulay,
\newblock J. Comput. Chem. {\bf 3}, 556 (1982).

\bibitem{ZAskarpour25}
Z.~Askarpour, M.~Nottoli, and B.~Stamm,
\newblock Journal of Chemical Theory and Computation {\bf 21}, 1643 (2025),
\newblock PMID: 39916515.

\bibitem{fernandez-alberti_identification_2012}
S.~Fernandez-Alberti, A.~E. Roitberg, T.~Nelson, and S.~Tretiak,
\newblock The Journal of Chemical Physics {\bf 137}, 014512 (2012).

\bibitem{meek_evaluation_2014}
G.~A. Meek and B.~G. Levine,
\newblock The Journal of Physical Chemistry Letters {\bf 5}, 2351 (2014).

\bibitem{wang_simple_2014}
L.~Wang and O.~V. Prezhdo,
\newblock The Journal of Physical Chemistry Letters {\bf 5}, 713 (2014).

\bibitem{burton_energy_2022}
H.~G.~A. Burton,
\newblock Journal of Chemical Theory and Computation {\bf 18}, 1512 (2022).

\bibitem{kowalczyk_self-consistent_2016}
T.~Kowalczyk, K.~Le, and S.~Irle,
\newblock Journal of Chemical Theory and Computation {\bf 12}, 313 (2016).

\bibitem{marie_excited_2023}
A.~Marie and H.~G.~A. Burton,
\newblock The Journal of Physical Chemistry A {\bf 127}, 4538 (2023).

\bibitem{zaitsevskii_discontinuities_1994}
A.~Zaitsevskii and J.-P. Malrieu,
\newblock Chemical Physics Letters {\bf 228}, 458 (1994).

\bibitem{saade_excited_2024}
S.~Saade and H.~G.~A. Burton,
\newblock Journal of Chemical Theory and Computation {\bf 20}, 5105 (2024).

\bibitem{tran_tracking_2019}
L.~N. Tran, J.~A.~R. Shea, and E.~Neuscamman,
\newblock Journal of Chemical Theory and Computation {\bf 15}, 4790 (2019).

\bibitem{Zhanserkeev21}
A.~Zhanserkeev, J.~J. Talbot, and R.~P. Steele,
\newblock J. Chem. Theory Comput. {\bf 17}, 4675 (2021).

\end{thebibliography}
\bibliographystyle{aip}

\end{document}